\documentclass[twocolumn,preprintnumbers,amsmath,amssymb]{revtex4-1} 

\usepackage[utf8]{inputenc}
\usepackage{graphics,graphicx}
\usepackage{amssymb}
\usepackage{amsmath}
\usepackage{upgreek}
\usepackage{bm}

\usepackage[usenames,dvipsnames,svgnames,table]{xcolor}

\usepackage[normalem]{ulem}

\def\braket#1{\mathinner{\langle{#1}\rangle}}	

\def\ab{\alpha_{2}}
\def\ac{\alpha_{3}}
\def\avLN{\braket{L(N)}}

\def\avnN{\braket{n_3(N)}}

\def\avON{\braket{O(N)}}
\def\avpcl{\braket{p_{\mathrm{c}}(\ell)}}
\def\avpclN{\braket{p_{\mathrm{c}}(\ell,N)}}
\def\avRgN{\braket{R_{\mathrm{g}}^{2}(N)}}

\def\avRl{\braket{R^{2}(\ell)}}
\def\avRlN{\braket{R^{2}(\ell,N)}}

\def\dlcenter{\delta \ell_{\text{center}}}
\def\dlcentermax{\delta \ell_{\text{center}}^{\text{max}}}
\def\dlmaxroot{\delta \ell^{\text{max}}_{\text{root}}}
\def\e{\mathrm{e}}
\def\G{\mathcal{G}}
\def\Hid{\mathcal{H}_{\text{id}}}
\def\Hint{\mathcal{H}_{\text{int}}}
\def\H{\mathcal{H}}
\def\kBT{k_{\text{B}}T}
\def\lK{\ell_{\text{K}}}
\def\Lmax{L_{\text{max}}}
\def\mubr{\mu_{\text{br}}}
\def\Nbr{N_{\text{br}}}
\def\Ncenter{N_{\text{center}}}
\def\Nmin{N_{\text{min}}}
\def\nupath{\nu_{\text{path}}}
\def\nuref{\nu_{\text{ref}}}

\def\pl{p_{N}(\ell)}
\def\prbl{p_{N}(\vec r|\ell)}
\def\prb{p_{N}(\vec r)}
\def\prl{p_{N}(\vec r|\ell)}
\def\pr{p_{N}(\vec r)}
\def\Q{\mathcal{Q}}
\def\redchi{\tilde{\chi}^2}
\def\Rg{R_{\mathrm{g}}}

\def\thetal{\theta_{\ell}}
\def\thetapath{\theta_{\text{path}}}
\def\thetatree{\theta_{\text{tree}}}

\def\tl{t_{\ell}}

\def\tMC{t_{\text{MC}}}
\def\tpath{t_{\text{path}}}
\def\ttree{t_{\text{tree}}}

\begin{document}


\title{Randomly branching $\uptheta$-polymers in two and three dimensions: \\ Average properties and distribution functions}

\author{Irene Adroher-Benítez}
\email[]{iadroher@sissa.it}
\affiliation{SISSA - Scuola Internazionale Superiore di Studi Avanzati, Via Bonomea 265, 34136 Trieste, Italy}
\author{Angelo Rosa}
\email[]{anrosa@sissa.it}
\affiliation{SISSA - Scuola Internazionale Superiore di Studi Avanzati, Via Bonomea 265, 34136 Trieste, Italy}

\date{\today}

\begin{abstract}
Motivated by renewed interest in the physics of branched polymers, we present here a complete characterization of the connectivity and spatial properties of $2$ and $3$-dimensional single-chain conformations of randomly branching polymers in $\uptheta$-solvent conditions obtained by Monte Carlo computer simulations.
The first part of the work focuses on polymer average properties, like the average polymer spatial size as a function of the total tree mass and the typical length of the average path length on the polymer backbone.
In the second part, we move beyond average chain behavior and we discuss the complete distribution functions for tree paths and tree spatial distances, which are shown to obey the classical Redner-des Cloizeaux functional form.
Our results were rationalized first by the systematic comparison to a Flory theory for branching polymers and, next, by generalized Fisher-Pincus relationships between scaling exponents of distribution functions.
For completeness, the properties of $\uptheta$-polymers were compared to their ideal ({\it i.e.}, no volume interactions) as well as good-solvent ({\it i.e.}, above the $\uptheta$-point) counterparts.
The results presented here brings to conclusion the recent work performed in our group [A. Rosa and R. Everaers, J. Phys. A: Math. Theor. 49, 345001 (2016), J. Chem. Phys. 145, 164906 (2016), Phys. Rev. E 95, 012117 (2017)] 
in the context of the scaling properties of branching polymers.
\end{abstract}


\maketitle

\section{Introduction}\label{sec:introduction}
Branched polymers or trees represent a fundamental class of polymers whose physics is far more intricate and range of applications far more wider than the familiar example of {\it linear} polymers~\cite{Burchard1999}. 

From the theoretical point of view, physical models for randomly branched polymers were introduced to account for the behavior of synthetic as well as biological macromolecules, ranging from
star molecules as in the classical work by Zimm and Stockmayer~\cite{ZimmStockmayer49} to more recent bio-oriented applications like the folding of single-stranded RNA molecules in viral capsids~\cite{YoffeGelbart2008,FangGelbart2011,GrosbergKellyBruinsma2017}
and the design of novel soft materials~\cite{CorsiCaponePCCP2019} for specific practical scopes as efficient drug delivery~\cite{MovellanVicent2015}.

In the more or less close past, there have been considerable efforts~\cite{vonFerberBlumenJCP2002,GurtovenkoBlumen2005,DolgushevBlumenMacromolThSimuls2011,Read2013,VargasLaraDouglas2018} in trying to predict how the effects of branching on polymer structure are expected to impact on polymer relaxation and dynamics.
In particular, our renewed interest in the field of branched polymers has been motivated because of the connection between those and the large scale behavior of unlinked and unknotted ring polymers in concentrated solutions and melt~\cite{KhokhlovNechaev1985,RubinsteinRingDynamicsPRL1986,ObukhovRubinsteinDukePRL1994,RosaEveraersPRL2014,GrosbergSoftMatter2014,RosaEveraersJCP2016} and in more generic topologically-constraining environments~\cite{RosaEveraersEPJE2019}.

From the theoretical point of view, {\it linear} polymers are described~\cite{DoiEdwardsBook,RubinsteinColbyBook} in terms of the expectation value of the square gyration radius $\avRgN$ of the chain which grows as a power law of the total weight of the polymer, $N$:
\begin{equation}\label{eq:Rg2vsN}
    \avRgN \sim N^{2\nu} \, .
\end{equation}
$\nu$ is the metric scaling exponent of the chain, and it depends on the spatial dimension $d$ and the nature of the solvent surrounding the polymer~\cite{RubinsteinColbyBook}:
in good solvents monomers effectively repel each other and $\nu \approx 3/(d+2)$ for $d\leq4$ which corresponds to the scaling exponent of the so-called self-avoiding random-walk~\cite{MadrasSokal1988,Sokal1994ArXiv,Sokal1996MC},
while in bad solvents monomer-monomer attraction is strong enough to fold the polymer into a compact globular state with $\nu=1/d$.
For many solvents, the quality depends on temperature $T$: at the so-called $\uptheta$-temperature $T = T_{\uptheta}$ repulsion and attraction balance each other almost exactly, thus the chain behaves like under quasi-ideal conditions~\cite{DesCloizeauxBook} with $\nu=1/2$ in $d=3$.

Because of their more complicate nature, a quantitative description of the structure of branched polymers needs the introduction of additional observables, namely~\cite{RosaEveraersJPhysA2016,RosaEveraersJCP2016}:
\begin{enumerate}
  \item The average path length, $\avLN$, between pairs of monomers on the tree as a function of $N$:
	\begin{equation}\label{eq:LvsN}
	  \avLN \sim N^{\rho} \, .
	\end{equation}
  \item The average branch weight, $\braket{\Nbr(N)}$, as a function of $N$: 
	\begin{equation}\label{eq:NbrvsN}
	  \braket{\Nbr(N)} \sim N^{\epsilon} \, .
	\end{equation}
  \item The mean square spatial distance, $\avRl$, between pairs of tree nodes as a function of their mutual path distance, $\ell$:
	\begin{equation}\label{eq:R2vsL}
	  \avRl \sim \ell^{2\nupath} \, .
	\end{equation}
  \item The mean contact probability, $\avpcl$, between pairs of tree nodes as a function of $\ell$:
	\begin{equation}\label{eq:PCvsL}
	  \avpcl \sim \ell^{-\nupath(d+\thetapath)} \, .
	\end{equation}
\end{enumerate}
Thus, the scaling exponent $\nu$ is complemented by other exponents $\rho$, $\epsilon$, $\nupath$ and $\thetapath$ which, once again, turn out to depend on spatial dimension $d$ and on the universality class of the system:
now, the latter depends not only on the quality of the surrounding solvent~\cite{RubinsteinColbyBook} as for the linear case, but it also depends on the chain connectivity being quenched or annealed~\cite{Gutin1993,CuiChenPRE1996,RosaEveraersJPhysA2016} and, in solutions of many chains, on the inter-chain polymer-polymer interactions~\cite{RosaEveraersJCP2016}.
Two additional relationships between the exponents complete the picture:
the obvious $\nu = \rho \, \nupath$ and the less trivial $\rho=\epsilon$~\cite{JansevanRensburgMadrasJPhysA1992}.

To our knowledge, only very few exact values of these exponents have been reported in the literature.
For ideal (\textit{i.e.} without excluded volume effects) tree polymers $\nupath=\rho=\epsilon=1/2=2\nu$ and $\thetapath=0$~\cite{ZimmStockmayer49,DaoudJoanny1981,RosaEveraersJPhysA2016},
while for $3d$ single trees in good solvent $\nu=1/2$~\cite{ParisiSourlasPRL1981}.
Otherwise, in the rest of the cases, approximate values for scaling exponents have been worked out by resorting to rather sophisticate numerical or theoretical tools, which the interested reader can be find reported and discussed in the recent review work~\cite{EveraersGrosbergRubinsteinRosaSoftMatter2017}.

By using a suitable combination of Flory theory, scaling arguments and numerical simulations, Rosa and Everaers contributed to characterize the physics of randomly branching polymers by providing predictions for the scaling exponents of Eqs.~\eqref{eq:Rg2vsN} to~\eqref{eq:PCvsL} for $3d$ single self-avoiding trees in good solvent with annealed and quenched branching statistics~\cite{RosaEveraersJPhysA2016} and for melts of trees in $2d$ and $3d$~\cite{RosaEveraersJCP2016}.
Later on, they extended~\cite{RosaEveraersPRE2017} the analysis of these simulations by studying the distribution functions for the different observables which contributed to highlight the limits of mean-field-like Flory theory in describing the structural properties of lattice trees~\cite{EveraersGrosbergRubinsteinRosaSoftMatter2017}.

In this article, we add the missing piece to the picture described in these previous works~\cite{RosaEveraersJPhysA2016,RosaEveraersJCP2016,RosaEveraersPRE2017} by considering the case of randomly branching polymers with annealed connectivity~\cite{Gutin1993,CuiChenPRE1996,RosaEveraersJPhysA2016} in $\uptheta$-solvent in $2d$ and $3d$.
In particular, we provide here a complete characterization of these specific polymer ensembles by studying their average properties in terms of the observables~\eqref{eq:Rg2vsN} to~\eqref{eq:PCvsL} and the associated distribution functions, and we compare those to corresponding ensembles of ideal trees and polymers in good solvent conditions.
As in works~\cite{RosaEveraersJPhysA2016,RosaEveraersJCP2016}, the discussion is guided through the systematic comparison to the predictions (and limitations) of the mean-field-like Flory theory~\cite{EveraersGrosbergRubinsteinRosaSoftMatter2017}.

The paper is organized as follows.
In Section \ref{sec:Theory} we briefly review the Flory theory for lattice trees, as well as the scaling properties of distribution functions that provide information beyond the mean-field approximation.
In Sec.~\ref{sec:methods} we describe the numerical model for the branching polymers, in particular the procedure to derive an adequate force field for the $\uptheta$-polymers, then the algorithm used to perform the Monte Carlo simulations and the methods employed to analyze the output data.
In Sec.~\ref{sec:results} we discuss the main results obtained in this work, while additional plots and Tables are placed in the Appendices at the end of the work.
Finally, in Sec.~\ref{sec:conclusions} we outline the conclusions.

\section{Theoretical background}\label{sec:Theory}

\subsection{Flory theory for randomly branching polymers}\label{sec:FloryTheory}
In spite of being based on rather crude assumptions, Flory theories~\cite{Flory1969,Isaacson1980,MaritanGiacometti2013,EveraersGrosbergRubinsteinRosaSoftMatter2017} provide remarkably accurate results for the scaling exponents of observables Eqs.~\eqref{eq:Rg2vsN}-\eqref{eq:R2vsL} (however (!) no concrete prediction for $\theta_{\rm path}$ aside from the trivial one can be formulated, as discussed below).
In this Section, we derive these theoretical exponents for interacting polymers in good and $\uptheta$-solvent conditions.
In this way, we are able to explore the differences between the scaling exponents obtained from the simulations and those expected from the mean-field Flory theory.

For randomly branching polymers in a generic solvent in $d$-dimensions 
and neglecting numerical prefactors, the Flory free energy ($\mathcal F$) is formulated as a balance of three terms:
\begin{equation}\label{eq:FloryFreeEn}
\frac{\mathcal F}{k_B T} \sim \frac{R^2}{\ell_K L} + \frac{L^2}{\ell_K^2 N} + v_p \frac{N^p}{R^{(p-1)d}} \, .
\end{equation}
The first and second terms account for the entropic contributions coming from stretching the polymer~\cite{FloryChemBook} between any two ends and from branching statistics~\cite{DeGennes1968,GrosbergNechaev2015}, respectively.
The third term is the expression for the $p$-body interaction which characterizes solvent-mediated interactions between monomers~\cite{MaritanGiacometti2013,EveraersGrosbergRubinsteinRosaSoftMatter2017}.
Thus, $p=2$ and $p=3$ describe good and $\uptheta$-solvent, respectively.

Minimization of Eq.~\eqref{eq:FloryFreeEn} with respect to $L$ and $R$ leads to:
\begin{eqnarray}
L & \sim & (\ell_K R^2 N)^{1/3} \, , \label{eq:MinizingFlory-L}\\
R & \sim & \ell_K N^{(3p+1)/(4+3(p-1)d)} \, , \label{eq:MinizingFlory-R}
\end{eqnarray}
which imply the following results for the scaling exponents:
\begin{eqnarray}
\nu  & = & \frac{3p+1}{4+3(p-1)d}, \\
\rho  & = & \epsilon = \frac{2(p+1)+(p-1)d}{4+3(p-1)d}, \\
\nupath & = & \frac{\nu}{\rho} = \frac{3p+1}{2(p+1)+(p-1)d}.
\label{eq:FloryExps}
\end{eqnarray}
For single randomly branching polymers in good solvent $p=2$, so Eqs.~\eqref{eq:FloryExps} give:
\begin{eqnarray}
\nu & = & \frac{7}{4+3d}, \\
\rho & = & \epsilon = \frac{6+d}{4+3d}, \\
\nupath & = & \frac{\nu}{\rho} = \frac{7}{6+d},
\label{eq:FloryExps-GoodSolvent}
\end{eqnarray}
while in $\uptheta$-solvent $p=3$ and
\begin{eqnarray}
\nu & = & \frac{5}{2+3d}, \\
\rho & = & \epsilon = \frac{4+d}{2+3d}, \\
\nupath & = & \frac{\nu}{\rho} = \frac{5}{4+d}.
\label{eq:FloryExps-ThetaSolvent}
\end{eqnarray}
Finally, because Flory theories neglect chain correlations {\it by construction}, $\thetapath=0$.

In this work we will mainly focus on $2d$ and $3d$ $\uptheta$-polymers and we will compare the predictions of Eqs.~\eqref{eq:FloryExps-ThetaSolvent} with the results of Monte Carlo computer simulations. For completeness, we discuss those together with the properties of ideal as well as randomly branching polymers in good solvent (Eqs.~\eqref{eq:FloryExps-GoodSolvent}).

\subsection{Beyond Flory theory: Distribution functions}\label{sec:TheoryDistributions}
Even though Flory theory provides a useful framework to study the scaling properties of branching polymers, its mean-field-like nature is based on rather crude assumptions~\cite{DeGennes1979,DesCloizeaux1990},
{\it in primis} the Gaussian functional form for quantifying the entropy in Eq.~(\ref{eq:FloryFreeEn}).

An example of the limitations of the Flory theory is given by the well known example of the end-to-end distribution function, $p(\vec r_{\rm ee})$, for $d$-dimensional self-avoiding {\it linear} ({\it i.e.}, unbranched) polymers in good solvent conditions.
It turns out in fact that, in the large ($N\rightarrow\infty$) polymerization limit $p(\vec r_{\rm ee})$ obeys the following scaling ansatz:
\begin{equation}\label{eq:SAWend2end}
p(\vec r_{\rm ee}) = \frac{1}{\langle R^2(N) \rangle^{d/2}} \, q \left( \frac{ \vec r_{\rm ee} } { \langle R^2(N) \rangle^{d/2} } \right) \, ,
\end{equation}
where $\langle R^2(N) \rangle \sim N^{2\nu}$ and
\begin{equation}\label{eq:RdC-q}
    q(x) = C \, x^{\theta} \exp\left[-\left( Kx \right)^{t}\right]
\end{equation}
satisfies the so-called Redner-des Cloizeaux (RdC) functional form~\cite{DesCloizeauxPRA1974,Redner1980,CaraccioloPelissettoJPA2011}.
The two constants $C$ and $K$ can be computed by imposing that:
(1)
$q(x)$ normalizes to 1
and
(2)
the second moment constitutes the only scaling length~\cite{EveraersJPhysA1995,RosaEveraersPRE2017}.
With these constraints we get easily the following analytical expressions:
\begin{eqnarray}
C &=& t \, \frac{\Gamma(1+\frac d2)\Gamma^{\frac{d+\theta}2}(\frac{2+d+\theta}t)}{d\,\pi^{d/2}\,\Gamma^{\frac{2+d+\theta}2}(\frac{d+\theta}t)} \, , \label{eq:RdC_C} \\
K^2 &=& \frac{\Gamma(\frac{2+d+\theta}t)}{\Gamma(\frac{d+\theta}t)} \, . \label{eq:RdC_K}
\end{eqnarray}
Here, $\Gamma=\Gamma(x)$ is the standard Euler's $\Gamma$-function.
Thus, the knowledge of the pair of exponents $(\theta,t)$ is enough to reconstruct the full distribution function in the asymptotic limit of large $N$.
Interestingly, the two exponents $\theta$ and $t$ are not completely independent and are related to the other scaling exponents.
Thus,
(a)
the ``mechanical'' Fisher-Pincus~\cite{Fisher1996,Pincus1976} relationship imposes that:
\begin{equation}\label{eq:SAWFisherPincus-t}
t = \frac{1}{1-\nu} \, ,
\end{equation}
while
(b):
\begin{equation}\label{eq:SAWFisherPincus-theta}
\theta = \frac{\gamma-1}{\nu}
\end{equation}
constitutes a sort of measure of the ``entropy'' of the walks since the exponent $\gamma$ appears~\cite{DuplantierJStatPhys1989,DesCloizeaux1990} in the partition function ${\mathcal Z}_N \sim N^{\gamma-1} \mu^N$ of the walks.
Notice that, with $\nu=1/2$ and $\gamma=1$ (or, $\theta=0$) the RdC function reduces to the classical Gaussian function describing ideal polymers~\cite{DoiEdwardsBook,RubinsteinColbyBook}.

As shown in the recent work~\cite{RosaEveraersPRE2017} by our group, the RdC formalism can be easily generalized to describe the scaling behavior of the distribution functions of the observables considered in Eqs.~\eqref{eq:Rg2vsN}-\eqref{eq:PCvsL}. 

To fix the ideas, we focus on the following functions:
(1)
the distribution function, $\pl$, of linear paths of contour length $\ell$ on polymers of weight $N$;
(2)
the distribution function, $\prl$, of the end-to-end spatial distances for linear paths of contour length $\ell$, and 
(3)
the distribution function of spatial distances between pairs of nodes, $\pr$.
The three functions are not independent, as they satisfy the obvious convolution-like identity:
\begin{equation}
\pr = \int_0^\infty \prl \, \pl \mathrm{d}\vec \ell \, .
\label{eq:Convolution}
\end{equation}

Asymptotically, these functions display universal behaviors and, respectively, can be expressed by the following scaling forms:
\begin{eqnarray}
    \pl & =   & \frac{1}{\avLN} \, q \left( \frac{\ell}{\avLN} \right) \, ,                                            \label{eq:RdC-pl} \\
    \prbl & = & \frac{1}{\avRlN^{d/2}} \, q \left( \frac{|\vec r|}{\avRlN^{1/2}} \right) \, ,                           \label{eq:RdC-prl} \\
    \prb & =  & \frac{1}{\left(2\avRgN\right)^{d/2}} \, q \left( \frac{|\vec r|}{\left(2\avRgN\right)^{1/2}} \right) \, . \label{eq:RdC-pr}
\end{eqnarray}
In all cases described by Eqs.~\eqref{eq:RdC-pl}-\eqref{eq:RdC-pr}, the function $q(x)$ obeys the RdC functional form Eq.~(\ref{eq:RdC-q}) with {\it novel} pairs of exponents called, respectively, $(\thetal, \tl)$, $(\thetapath,\tpath)$ and $(\thetatree, \ttree)$~\cite{RosaEveraersPRE2017},
whose knowledge appears of fundamental importance for understanding the physics of branching polymers in different solvent conditions.

As for the functions in Eqs.~(\ref{eq:RdC-prl}) and~(\ref{eq:RdC-pr}), corresponding constants $C$ and $K$ are given by expressions analogous to Eqs.~(\ref{eq:RdC_C}) and~(\ref{eq:RdC_K}).
As for Eq.~(\ref{eq:RdC-pl}) instead, the obvious normalization to $1$ of $q(x)$ is accompanied by imposing that the {\it first} (rather than the second~\cite{RosaEveraersPRE2017}) moment constitutes the only scaling length, thus implying the following expressions:
\begin{eqnarray}
C &=& \tl \, \frac{\Gamma^{\thetal+1}((\thetal+2)/\tl)}{\Gamma^{\thetal+2}((\thetal+1)/\tl)} \, , \label{eq:RdC_C_l} \\
K &=& \frac{\Gamma((\thetal+2)/\tl)}{\Gamma((\thetal+1)/\tl)} \, . \label{eq:RdC_K_l}
\end{eqnarray}

In analogy to the example of self-avoiding linear polymers, the different pairs of exponents $(\theta, t)$ are quantitatively related to the exponents $(\nu, \rho, \nupath)$ characterizing the scaling behavior of the trees average properties through generalized Fisher-Pincus~\cite{Fisher1996,Pincus1976} relations, as documented first in Ref.~\cite{RosaEveraersPRE2017}.
Thus the exponent $\rho$, which describes the scaling of the average path length, can be used to compute the exponents $(\thetal,\tl)$ describing the distribution of linear paths of contour length, $\pl$,
\begin{eqnarray}
  \thetal & = & \frac{1}{\rho} - 1 \, , \label{eq:FP-thetal} \\
  \tl  & = & \frac{1}{1-\rho} \, . \label{eq:FP-tl}
\end{eqnarray}
Similarly, the exponent $\tpath$ is related to $\nupath$:
\begin{equation}\label{eq:FP-tpath}
  \tpath = \frac{1}{1-\nupath} \, ,
\end{equation}
while, interestingly, $\thetapath$ (which describes the decay of the mean contact probablity, Eq.~(\ref{eq:PCvsL})) is independent by the others and, then, it constitutes a genuinely {\it novel} exponent.
Finally, the exponents for the distribution of spatial distances between pairs of nodes, $\pr$, are related to the the metric scaling exponent of the polymer chain $\nu$ and $\theta_{\rm path}$ by:
\begin{eqnarray}
  \thetatree & = & \min\left(\thetapath, \frac{1}{\nu} - d\right) \, , \label{eq:FP-thetatree} \\
  \ttree  & = & \frac{1}{1-\nu} \, . \label{eq:FP-ttree}
\end{eqnarray}
The interested reader can find a complete account on the mathematical derivation and physical meaning of these relations in Ref.~\cite{RosaEveraersPRE2017}.

\section{Model and methods}\label{sec:methods}
In this article we generalize the polymer model and Monte Carlo algorithm described in the former works~\cite{RosaEveraersJPhysA2016,RosaEveraersJCP2016},
the details relevant here being summarized respectively in Sec.~\ref{sec:BranchinPolymsSqCubicLattices} and Sec.~\ref{sec:MCsimuls}.
The main novelty of this work, namely how to model $\uptheta$-polymers, is explained in detail in Sec.~\ref{sec:ThetaInteraction}.
Finally, notation, algorithms and methods (also inspired by works~\cite{RosaEveraersJPhysA2016,RosaEveraersJCP2016}) employed in the characterization of polymer conformations and estimation of scaling exponents are described in Sec.~\ref{sec:NumMethodsScalingExps}.

\subsection{Branching polymers on the square and cubic lattices}\label{sec:BranchinPolymsSqCubicLattices}

%
\begin{figure}[htb!]
  \includegraphics[width=0.9\linewidth]{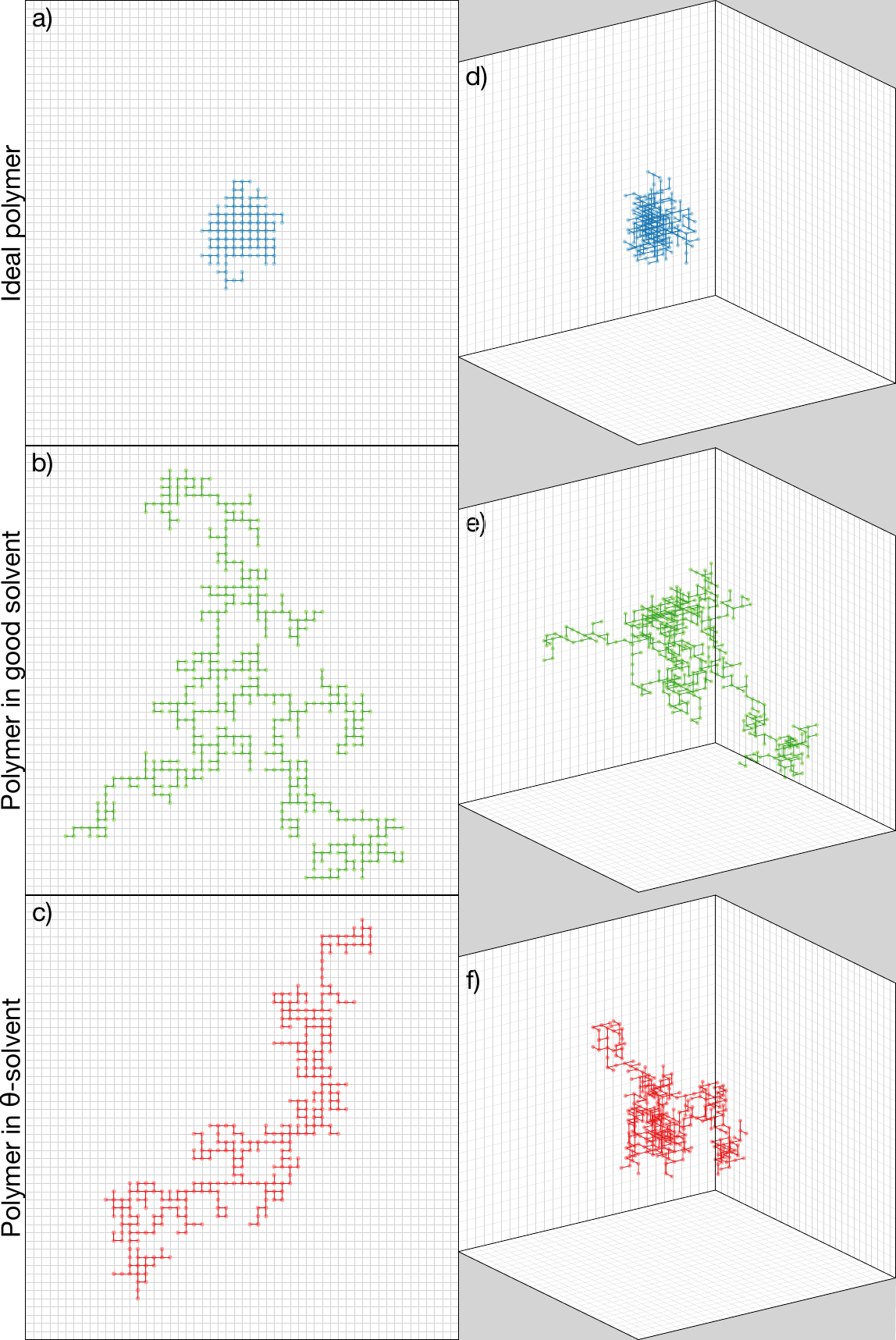}
  \caption{
  Representative conformations for $2d$ (l.h.s. panels) and $3d$ (r.h.s. panels) randomly branching polymers from different ensembles:
  (a, d) ideal polymers,
  (b, e) single randomly branching polymers in good solvent,
  (c, f) single randomly branching polymers in $\uptheta$-solvent.
  For interacting polymers, the size of the simulation box is typically much larger than the lattice portion shown here so to guarantee the dilute regime.
  }
  \label{fig:structures}
\end{figure}

We model randomly branching polymers as {\it loop-free} trees on the square ($d=2$) and cubic ($d=3$) lattices 
with periodic boundary conditions. 
For self-interacting polymers (see Sec.~\ref{sec:MCsimuls}) the simulation box was chosen large enough to ensure that we have been working in the dilute regime.

To fix the notation, we consider polymers or trees consisting of a branched structure in which $N+1$ nodes are connected by $N$ Kuhn~\cite{RubinsteinColbyBook} segments of unit length $\lK$ and unit mass $m$.
For simplicity and without any loss of generality~\cite{RosaEveraersJPhysA2016,RosaEveraersJCP2016} we limit the maximal functionality of each node (corresponding to the number of bonds protruding from the node) to $f\leq3$. 

We stress that we consider trees with {\it annealed} connectivity~\cite{RosaEveraersJPhysA2016}, which means that the location of the branching points undergoes thermal fluctuations as the result of the coupling to an external control parameter.
It turns out that this is very different from the ensemble where connectivity is kept {\it quenched} as in the case of chemically-synthesized polymers,
{\it branching} polymers with annealed connectivity and {\it branched} polymers with quenched connectivity belonging to different universality classes~\cite{Gutin1993,CuiChenPRE1996,RosaEveraersJPhysA2016}.
Single representative conformations for each of the different polymer ensembles considered in this work are illustrated in Figure~\ref{fig:structures}.

\subsection{Monte Carlo computer simulations}\label{sec:MCsimuls}

\subsubsection{The algorithm}\label{sec:AmoebaAlgo}

%
\begin{figure}[htb!]
  \includegraphics[width=0.75\linewidth]{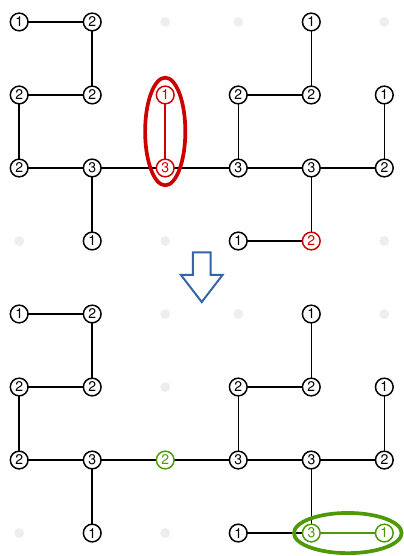}
  \caption{
  Schematic drawing describing the version of the amoeba algorithm~\cite{Seitz1981} employed here.
  Each node of the tree is identified by its corresponding functionality, $f$.
  (Top)
  The red nodes indicate the leaf chosen to be cut and the node were it will be pasted.
  (Bottom)
  The new polymer configuration obtained, with corresponding updated links and nodes functionalities.
  }
  \label{fig:algorithm}
\end{figure}

Monte Carlo simulations of randomly branching polymers are performed according to a slightly modified version of the so-called ``amoeba'' algorithm by Seitz and Klein~\cite{Seitz1981}.
In this algorithm, each new configuration is generated from the previous one by randomly cutting a leaf from the tree and then reconnecting it randomly to one of the other nodes with functionaly $f\leq2$, thus constraining the single-node functionality to be not larger than $f=3$ (see Sec.~\ref{sec:BranchinPolymsSqCubicLattices}).
A schematic example of this procedure is shown in Fig.~\ref{fig:algorithm}.

Each polymer configuration is characterized by the position of all its nodes, $\Gamma = \{ \vec{r}_{1}, ..., \vec{r}_{N+1} \}$, and their connectivity, $\G$.
Both $\Gamma$ and $\G$ are modified in a trial move of the amoeba algorithm.
This move will be accepted with probability given by the standard Metropolis~\cite{KrauthBook} algorithm accounting for detailed balance:
\begin{equation}
\text{acc}_{i \to f} = \text{min}\left\{ 1, \frac{n_{1}(i)}{n_{1}(f)}\e^{-\frac{1}{\kBT}[\H(\Gamma_{f},\G_{f})-\H(\Gamma_{i},\G_{i})]} \right\} \, ,
\label{eq:Metropolis}
\end{equation}
where $1/\kBT$ is the Boltzmann factor, $n_{1}$ is the (initial and final) total number of nodes with functionality $f=1$ and $\H(\Gamma,\G)$ is the total interaction Hamiltonian (described in Sec.~\ref{sec:ThetaInteraction}) between nodes.

\subsubsection{Polymer equilibration and statistics}\label{sec:PolymerEquilibration}
Single tree conformations are initially prepared as ideal random walks on the lattice. 
The equilibration of our systems was checked by monitoring (see Fig.~\ref{fig:Rg2-n3-tMC} in the Appendix) that
(1) the ensemble-average square gyration radius of the polymer, $\langle R_{\rm g}^2(t_{\rm MC}) \rangle$,
and
(2) the ensemble-average number of branching nodes, $\langle n_3(t_{\rm MC}) \rangle$,
as functions of the Monte Carlo ``time'' steps $t_{\rm MC}$ both reach equilibrium values.
The total number, $M$, of statistically independent tree conformations used for the averages is:
$M=100$ for $3 \leq N \leq 10$;
$M=1000$ for $20 \leq N \leq 900$;
$M=2000$ for $N=1800$.

\subsection{Interaction Hamiltonian for branching polymers}\label{sec:ThetaInteraction}
The total energy of the system is given by the sum of two contributions, one ideal and one due to the interactions between nodes~\cite{RosaEveraersJPhysA2016,RosaEveraersJCP2016}:
\begin{equation}
  \H(\Gamma,\G) = \Hid(\G) + \Hint(\Gamma) \, .
  \label{eq:Htotal}
\end{equation}

The ideal contribution, $\Hid(\G)$, controls the connectivity of the chain and it is expressed in terms of the coupling between the chemical potential of branching points, $\mubr$, and the total number of 3-functional (branching) nodes in the polymer, $n_{3}(\G)$:
\begin{equation}\label{eq:Hid}
  \frac{\Hid (\G)}{\kBT} = \mubr \, n_{3}(\G) \, .
\end{equation}
The value $\mubr$ was fixed to be equal to $-2.0$.
For ideal polymers, this implies an average fraction of branching points $\approx 0.4$, see Refs.~\cite{RosaEveraersJPhysA2016,RosaEveraersJCP2016} and Fig.~\ref{fig:n3-N}.

The interaction term, $\Hint(\Gamma)$, between tree nodes is described as the sum of $2$- and $3$-body interactions,
\begin{equation}\label{eq:Hint}
  \frac{\Hint(\Gamma)}{\kBT} = \ab \displaystyle\sum_{j\in \text{lattice sites}} \kappa_{j}^{2} + \ac \displaystyle\sum_{j\in \text{lattice sites}} \kappa_{j}^{3} \, ,
\end{equation}
were $\kappa_{j}$ is the total number of Kuhn segments inside the elementary cell centered at the lattice site $j$.
An expression like Eq.~\eqref{eq:Hint} with $\ab>0$ and $\ac=0$ was already considered in Refs.~\cite{RosaEveraersJPhysA2016,RosaEveraersJCP2016} for modelling single branching polymers in good solvent or branching polymers in melt, as net $2$-body repulsive interactions are known to dominate polymer behavior in these regimes~\cite{MaritanGiacometti2013,EveraersGrosbergRubinsteinRosaSoftMatter2017}.

For polymers at the $\uptheta$-point, repulsive interactions are compensated by a monomer-monomer attraction at short spatial separations.
For {\it linear} polymers this has the important implications that their behavior is quasi-ideal, {\it i.e.} the scaling exponent $\nu \approx 1/2$~\cite{RubinsteinColbyBook}.
To properly model branching polymers in $\uptheta$-conditions we have then balanced the two terms in Eq.~\eqref{eq:Hint}:
an attractive 2-body interaction ($\ab<0$) is needed in order to overcome the volume exclusion,
which is taken into account through the repulsive 3-body interaction ($\ac>0$). 

\begin{figure}[htb!]
  \includegraphics[width=0.8\linewidth]{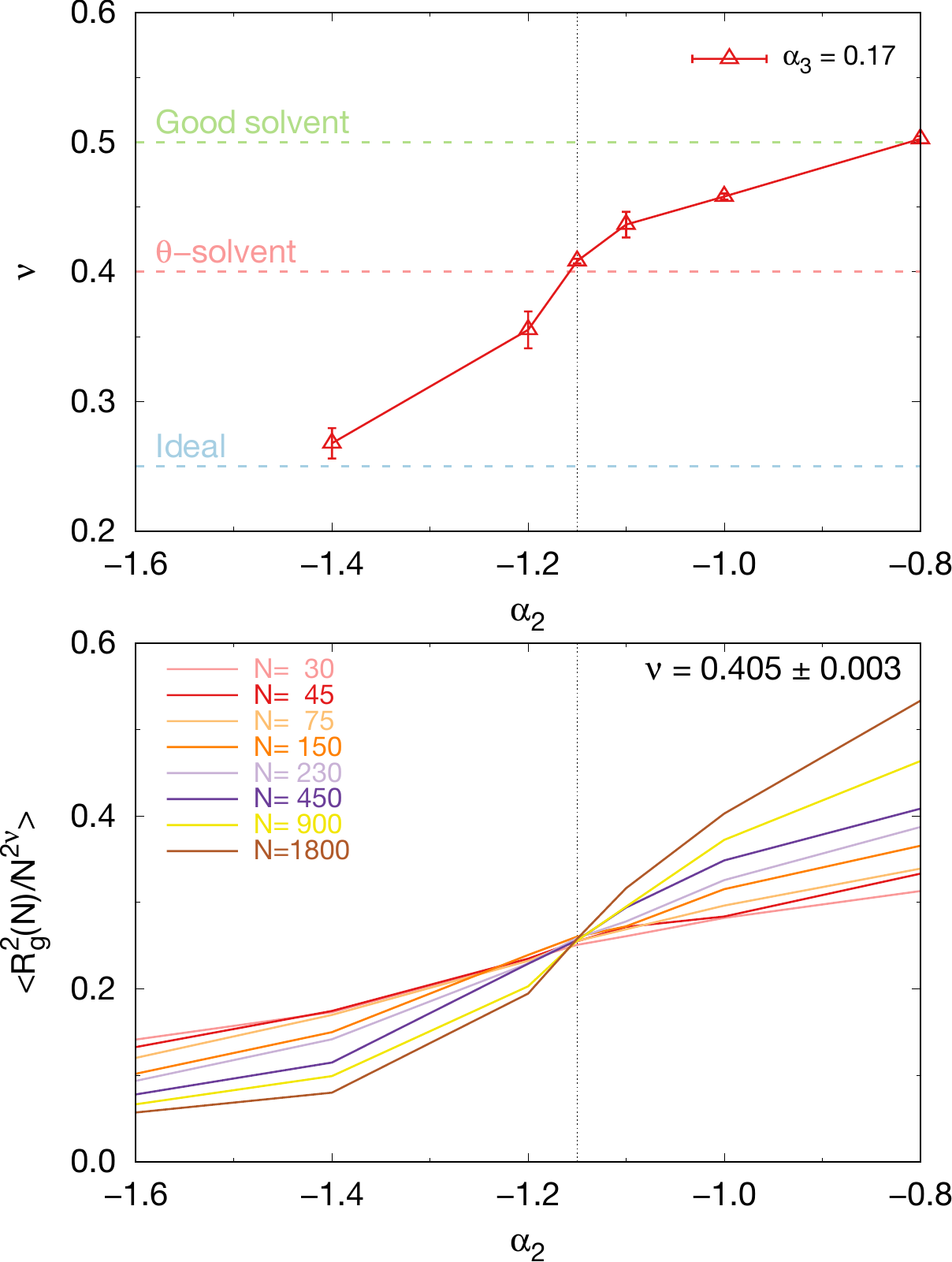}
  \caption{
  (Top)
  Scaling exponent $\nu$ for $3d$ branching polymers calculated for different values of the $2$-body interaction parameter, $\ab$, with given 3-body interaction parameter $\ac = 0.17$.
  $\uptheta$-solvent conditions correspond to the value $\ab=-1.15$ where the red curve intercepts the straight line (in red) $\nu=0.405\pm0.003$.
  (Bottom)
  Corresponding normalized mean square gyration radii, $\avRgN/N^{2\nu}$, with $\nu=0.405\pm0.003$. 
  }
  \label{fig:interactions}
\end{figure}

Finding the adequate values for $\ab$ and $\ac$ was done in the following way, as schematically illustrated in Fig.~\ref{fig:interactions} (top).
We considered $3d$ branching polymers and we found that for the pair ($\ab, \ac) = (0.0, 0.17)$ we observe the expected scaling behavior of branching polymers in good solvent conditions, $\avRgN \sim N^{2\nu} \sim N^1$~\cite{ParisiSourlasPRL1981}.
Then, at fixed $\ac$, we decreased progressively $\ab < 0$ and measured the corresponding scaling exponent $\nu$, which diminishes accordingly.
At the value $\ab=-1.15$ we found that the exponent is equal to $\nu=0.405\pm0.003$, which is in very good agreement with the most accurate reference estimate $\nuref = 0.400 \pm 0.005$ obtained by Madras and Janse van Rensburg from Monte Carlo computer simulations of $3$-dimensional $\uptheta$-trees~\cite{Madras1997,EveraersGrosbergRubinsteinRosaSoftMatter2017}. 
For even lower values of $\ab$, the polymers are found to collapse into quasi-ideal conformations with $\nu\approx1/4$.
Fig.~\ref{fig:interactions} (bottom) illustrates nicely the effectiveness of this procedure by plotting the quantity $\avRgN / N^{2\nu}$ with $\nu=0.405\pm0.003$ as a function of $\ab$ and for different $N$.
For $\ab=-1.15$ all curves intersect each other at the same point where finite-$N$ effects appear negligible.
Conversely, for smaller (respectively, larger) values of $\ac$ the different curves show a decreasing (resp., increasing) trend at increasing values of $N$.

Importantly, while the value of $\nu$ for our three-dimensional $\uptheta$-trees was obtained by fitting the interaction parameters of Eq.~\eqref{eq:Hint} so to reproduce the numerical result by Madras and Janse van Rensburg, we anticipate (see Sec.~\ref{sec:NumericalResults}) that the same parameters can be employed to model $\uptheta$-trees in two dimensions without additional fine tuning.
This finding demonstrates that our lattice model describes correctly the physics of randomly branching $\uptheta$-polymers.

\subsection{Analysis of polymer conformations}\label{sec:NumMethodsScalingExps}

%
\begin{table*}[htb!]
\caption{List of observables and distribution functions considered in this work, with description and reference to the corresponding figures.
These quantities have been used to characterize other trees ensembles, see Refs.~\cite{RosaEveraersJPhysA2016,RosaEveraersJCP2016,RosaEveraersPRE2017}.}
\begin{tabular}{ccc}
\hline
\hline
\\
Notation & Description & Figure \\
\hline
$\avnN$ & Average fraction of $3$-functional nodes as a function of the total tree weight $N$ & \ref{fig:n3-N} \\
$\avLN$ & Average path length between pairs of nodes as a function of the total tree weight $N$ & \ref{fig:L-dlcenter-dlcentermax-N} \\
$\braket{\dlcenter(N)}$ & Average path distance of nodes from the central node as a function of the total tree weight $N$ & \ref{fig:L-dlcenter-dlcentermax-N} \\
$\braket{\dlcentermax(N)}$ & Average {\it longest} path distance of nodes from the central node as a function of the total tree weight $N$ & \ref{fig:L-dlcenter-dlcentermax-N} \\
$\braket{\Nbr(\dlmaxroot)}$ & Average branch weight as a function of the longest path to the branch root $\dlmaxroot$ & \ref{fig:Nbr-dlroot-Ncenter-dlcenter} \\
$\braket{\Ncenter(\dlcenter)}$ & Average branch weight of paths whose distance from the tree center does not exceed $\dlcenter$ & \ref{fig:Nbr-dlroot-Ncenter-dlcenter} \\
$\braket{\Nbr(N)}$ & Average branch weight as a function of the total tree weight $N$ & \ref{fig:Nbr-N} \\
$\braket{R^{2}(\ell=\avLN)}$ & Mean-square end-to-end spatial distance for paths of contour length $\ell = \avLN$ & \ref{fig:R2-L-Lmax} \\
$\braket{R^{2}(\Lmax(N))}$ & Mean-square end-to-end spatial distance of the longest paths & \ref{fig:R2-L-Lmax} \\
$\avRlN$ & Mean-square end-to-end spatial distance for paths of contour length $\ell$ & \ref{fig:R2-pc-l} \\
$\avpclN$ & Mean closure probability for paths of contour length $\ell$ & \ref{fig:R2-pc-l} \\
$\avRgN$ & Mean-square tree gyration radius as a function of the total tree weight $N$ & \ref{fig:Rg2-N} \\
$\braket{\Lambda_{1,2,3}^{2}(N)}$ & Mean-square eigenvalues of the tree gyration tensor as a function of the total tree weight $N$ & \ref{fig:lambda-N} \\
$\pl$ & Probability distribution function of tree paths of total length $\ell$ & \ref{fig:pl-l} \\
$\prl$ & Probability distribution function of end-to-end vectors $\vec r$ between pairs of nodes of total path length $\ell$ & \ref{fig:prl-r} \\
$\pr$ & Probability distribution function of end-to-end vectors $\vec r$ between tree nodes & \ref{fig:pr-r} \\
\hline
\hline
\end{tabular}
\label{tab:ListOfObservables}
\end{table*}

Equilibrated polymer conformations obtained by Monte Carlo computer simulations were analyzed by following closely the definitions, algorithms and tools described in Refs.~\cite{RosaEveraersJPhysA2016,RosaEveraersJCP2016,RosaEveraersPRE2017}.
In particular, for characterizing the scaling behaviors of trees connectivity and spatial structure we adopt here the same terminology of these papers so, to avoid unnecessary repetition, we have recapitulated the complete list of observables in Table~\ref{tab:ListOfObservables}.
Then, in Sec.~\ref{sec:BurningAlg} we have briefly described the so-called ``burning'' algorithm necessary, in particular, to extract information on the observables quantifying trees connectivity.
Finally, scaling exponents and trees asymptotic properties are estimated by the finite-size scaling analysis presented in Sec.~\ref{sec:EstimatingScalExpsAvProps}.
The reader interested in more details and results concerning other polymer ensembles is invited to look into publications~\cite{RosaEveraersJPhysA2016,RosaEveraersJCP2016,RosaEveraersPRE2017}.

\subsubsection{Analysis of nodes connectivity via burning}\label{sec:BurningAlg}
In order to analyze the node-to-node connectivity we have resorted to a close variant of the ``burning'' algorithm originally proposed for percolation clusters~\cite{Herrmann1984}.
Essentially, this algorithm consists of two steps:
\begin{enumerate}
\item
An inward step, during which the polymer graph is ``burned'' from outside to inside by removal of all tips (the nodes with functionality $f=1$) in order to obtain a smaller tree to which the algorithm is then applied recursively.
The procedure stops when only one single node is left (the central node of the original tree).
\item
An outward step, consisting in the advancement from the center to the periphery.
\end{enumerate}
In the first step one collects information about the mass and shape of branches,
while in the second one reconstructs the distances of nodes from the centre of the tree.
The unique minimal path $\ell_{ij}$ between any pairs of tree nodes $i$ and $j$ is obtained by modifying the burning algorithm ensuring that it does not proceed inward of node $i$ and node $j$.
As our trees contain no loops {\it by construction}, the procedure ends with a single linear polymer of contour distance $=\ell_{ij}$.
The interested reader may find a more extensive illustration of the burning procedure in Ref.~\cite{RosaEveraersJPhysA2016}.

\subsubsection{Estimating scaling exponents: Average properties}\label{sec:EstimatingScalExpsAvProps}

%
\begin{table*}[htb!]
\caption{Final estimated for the scaling exponents of the lattice trees ensembles considered in this work.}
\begin{tabular}{cccccccccc}
\hline
\hline
\\
\multicolumn{10}{c}{2-dimensions}\\
\hline
& & \multicolumn{2}{c}{Ideal polymer} & & \multicolumn{2}{c}{Good solvent} & & \multicolumn{2}{c}{$\uptheta$-solvent} \\
\hline
& & Flory & Simulations & & Flory & Simulations & & Flory & Simulations \\ \hline
$\rho$ & & 0.5 & $0.48 \pm 0.06$ & & 0.8 & $0.739 \pm 0.016$ & & 0.75  & $0.711 \pm 0.016$ \\
$\epsilon$ & & 0.5 & $0.49 \pm 0.04$ & & 0.8 & $0.743 \pm 0.004$ & & 0.75  & $0.720 \pm 0.011$ \\
$\nu$ & & 0.25 & $0.24 \pm 0.07$ & & 0.7 & $0.624 \pm 0.005$ & & 0.625 & $0.61 \pm 0.10$ \\
$\nupath$  & & 0.5 & $0.51 \pm 0.02$ & & 0.875 & $0.836 \pm 0.009$ & & 0.833 & $0.88 \pm 0.04$ \\
$\thetapath$ & & - & $-0.30 \pm 0.14$ & & - & $1.6 \pm 0.2$ & & - & $1.6 \pm 0.3$ \\
\hline
\\
\multicolumn{10}{c}{3-dimensions}\\
\hline
& & \multicolumn{2}{c}{Ideal polymer\footnotemark[1]} & & \multicolumn{2}{c}{Good solvent\footnotemark[1]} & & \multicolumn{2}{c}{$\uptheta$-solvent} \\
\hline
& & Flory & Simulations & & Flory & Simulations & & Flory & Simulations \\
\hline
$\rho$ & & 0.50 & $0.49 \pm 0.04$ & & 0.692 & $0.64 \pm 0.02$ & & 0.636 & $0.585 \pm 0.018$ \\
$\epsilon$ & & 0.50 & $0.536 \pm 0.007$ & & 0.692 & $0.655 \pm 0.009$ & & 0.636 & $0.591 \pm 0.011$ \\
$\nu$ & & 0.25 & $0.25 \pm 0.02$ & & 0.538 & $0.48 \pm 0.04$ & & 0.455 & $0.405 \pm 0.003$ \\
$\nupath$ & & 0.50 & $0.509 \pm 0.009$ & & 0.778 & $0.74 \pm 0.02$ & & 0.714 & $0.686 \pm 0.016$ \\
$\thetapath$ & & - & $-0.04 \pm 0.04$ & & - & $1.30 \pm 0.10$ & & - & $0.83 \pm 0.05$ \\
\hline
\hline
\end{tabular}
\footnotetext[1]{Results for these ensembles were discussed in Ref.~\cite{RosaEveraersJPhysA2016}. They are reshown here for the purpose of comparison.}
\label{tab:exponents}
\end{table*}

Accurate evaluation of scaling exponents for the different observables Eqs.~\eqref{eq:Rg2vsN}-\eqref{eq:PCvsL} is non-trivial, as numerical procedures are typically plagued by finite-$N$ effects~\cite{JansevanRensburgMadrasJPhysA1992}.
In order to overcome this issue, we resort to the numerical strategy employed in the former works~\cite{RosaEveraersJPhysA2016,RosaEveraersJCP2016,RosaEveraersPRE2017} dedicated to branching polymers.

Let $O$ be a generic observable which depends on polymer size $N$ so that $\avON \sim N^\gamma$ for large $N$ where $\gamma$ is the corresponding scaling exponent.
A first estimate of the exponent $\gamma = \gamma_1 \pm \delta\gamma_1$ with corresponding statistical error is obtained by best fit of $\log \avON$ {\it vs.} $\log N$ for $N\geq450$ to the straight line:
\begin{equation}
    \log\avON = a + \gamma_1 \log N \, .
    \label{eq:fit2par}
\end{equation}
Then, in order to account for finite-$N$ effects and thus estimate systematic errors 
we best fit the data for the full range $N\geq10$ to the modified function: 
\begin{equation}
    \log\avON = a + b N^{-\Delta} + \gamma_2 \log N \, ,
    \label{eq:correct-scaling-log}
\end{equation}
which contains a proper correction-to-scaling~\cite{JansevanRensburgMadrasJPhysA1992} term.
In practice, instead of solving the non-linear fit with parameters ($a$, $b$, $\Delta$ and $\gamma_2$), we have linearized Eq.~\eqref{eq:correct-scaling-log} around some $\Delta_0$
\begin{equation}
    \log\avON = a + b N^{-\Delta_{0}} + \gamma_2 \log N + b \, (\Delta_{0} - \Delta)N^{-\Delta_{0}}\log N \, ,
    \label{eq:fit4par} 
\end{equation}
and linear fit the data accordingly by using different $\Delta_0$ values~\cite{RosaEveraersJPhysA2016}.
Thus, the final estimate for the scaling exponent $\gamma = \gamma_2\pm\delta\gamma_2$ comes from the fit whose $\Delta_0$ value makes the term $N^{-\Delta_{0}}\log N$ vanishing.

The quality of both fit procedures Eqs.~\eqref{eq:fit2par} and~\eqref{eq:fit4par} is checked by means of standard statistical analysis~\cite{NumericalRecipes}.
The fit is deemed to be reliable when the normalized $\chi$-square, $\redchi \equiv \frac{\chi^2}{D-f}\approx 1$.
$\chi^2$ is calculated by minimizing the weighted square deviation between the data and the model, and $D - f$ is the number of degrees of freedom, calculated as the difference between the number of data points ($D$) and the number of fit parameters ($f$).
The corresponding $\mathcal Q(D-f, \chi^2)$-values provide a quantitative indicator for the likelihood that $\chi^2$ should exceed the observed value, if the model were correct~\cite{NumericalRecipes}.
The results of all fits are reported together with the corresponding errors, $\redchi$ and $\mathcal Q$ values in Tables \ref{tab:fit-scaling_rho-epsilon} and \ref{tab:fit-scaling_nu-nupath} in the Appendix.
The reader will notice that there are a few cases that required a separate analysis because the second fitting procedure could not be trusted in virtue of its poor performance in modeling the data.

Final estimation for $\gamma$ is given by
$\gamma = \displaystyle\frac{\gamma_1+\gamma_2}{2} \pm \sqrt{\delta\gamma_{\rm stat}^2 + \delta\gamma_{\rm syst}^2}$,
where
$\delta\gamma_{\mathrm{stat}} = \max(\delta\gamma_1, \delta\gamma_2)$ provides the estimate for the statistical error
and
$\delta\gamma_{\mathrm{syst}} = \displaystyle\frac{|\gamma_1-\gamma_2|}{2}$ accounts for the systematic error~\cite{RosaEveraersJPhysA2016}.
With this procedure we have obtained the values of the scaling exponents for the scaling relations \eqref{eq:Rg2vsN}-\eqref{eq:PCvsL} summarized in Table~\ref{tab:exponents}.

\subsubsection{Estimating scaling exponents: Distribution functions}\label{sec:EstimatingScalExpsDistrFuncts}

%
\begin{table*}[htb!]
\caption{Scaling exponents for distribution functions $\pl$ (Eq.~(\ref{eq:RdC-pl})), $\prl$ (Eq.~(\ref{eq:RdC-prl})) and $\pr$ (Eq.~(\ref{eq:RdC-pr})).
For each ensemble, the first and second columns contain the values calculated from Fisher-Pincus (FP) relations, Eqs.~(\ref{eq:FP-thetal})-(\ref{eq:FP-ttree}), after substitutions of the scaling exponents summarized in Table \ref{tab:exponents}:
(first column) exponents according to the Flory theory and (second column) exponents estimated from the large-scale behaviors of observables measured from computer simulations.
The third column lists the final estimations obtained by extrapolating the scaling exponents evaluated from the fits of the distribution functions.
}
\begin{tabular}{cccccccccccccccc}
\hline
\hline
\\
\multicolumn{16}{c}{2-dimensions} \\ 
\hline
& & & \multicolumn{3}{c}{Ideal polymer} & & & \multicolumn{3}{c}{Good solvent} & & & \multicolumn{3}{c}{$\uptheta$-solvent} \\
\hline
& & & FP & FP & Extrapolation & & & FP & FP & Extrapolation & & & FP & FP & Extrapolation \\
& & & Flory  & Simulations & Simulations & & & Flory & Simulations & Simulations & & & Flory & Simulations & Simulations \\
\hline
$\thetal$ & & & $1.00$ & $1.1 \pm 0.3$ & $0.96 \pm 0.06$ & & & $0.25$ &  $0.35 \pm 0.03$ & $0.294 \pm 0.001$ & & & $0.33$ & $0.41 \pm 0.03$ & $0.37 \pm 0.04$ \\
$\tl$ & & & $2.00$ & $1.9 \pm 0.2$ & $2.11 \pm 0.05$ & & & $5.00$ & $3.83 \pm 0.23$ & $2.48 \pm 0.03$ & & & $4.00$ & $3.46 \pm 0.19$ & $2.15 \pm 0.07$ \\
$\thetapath$ & & & $0$ & $0$ & $-0.01 \pm 0.03$ & & & $>0$ & $>0$ & $1.74 \pm 0.09$ & & & $>0$ & $>0$ & $1.2 \pm 0.1$ \\
$\tpath$ & & & $2.00$ & $2.04 \pm 0.08$ & $2.07 \pm 0.06$ & & & $8.00$ & $6.10 \pm 0.33$ & $7.5 \pm 0.2$ & & & $6.00$ & $8.33 \pm 2.78$ & $5.8 \pm 0.3$ \\
$\thetatree$ & & & $0.00$ & $-0.30 \pm 0.14$ & $0.02 \pm 0.005$ & & & $-0.57$ & $-0.397 \pm 0.013$ & $-0.457 \pm 0.001$ & & & $-0.40$ & $-0.36 \pm 0.269$ & $-0.379 \pm 0.003$ \\
$\ttree$ & & & $1.33$ &  $1.32 \pm 0.12$ & $1.34 \pm 0.01$ & & & $3.33$ & $2.66 \pm 0.04$ & $2.154 \pm 0.002$ & & & $2.67$ & $2.56 \pm 0.66$ & $1.867 \pm 0.003$ \\
\hline
\\
\multicolumn{16}{c}{3-dimensions}\\
\hline
& & & \multicolumn{3}{c}{Ideal polymer\footnotemark[2]} & & & \multicolumn{3}{c}{Good solvent\footnotemark[2]} & & & \multicolumn{3}{c}{$\uptheta$-solvent} \\
\hline             
& & & FP & FP & Extrapolation & & & FP & FP & Extrapolation & & & FP & FP & Extrapolation \\
& & & Flory & Simulations & Simulations & & & Flory & Simulations & Simulations & & & Flory & Simulations & Simulations \\
\hline
$\thetal$ & & & $1.00$ &  $1.04 \pm 0.17$ & $1.05 \pm 0.2$ & & & $0.44$ & $0.56 \pm 0.05$ & $0.53 \pm 0.02$ & & & $0.57$ & $0.71 \pm 0.05$ &  $0.68 \pm 0.01$ \\
$\tl$ & & & $2.00$ &  $1.96 \pm 0.15$ &  $2.02 \pm 0.1$ & & & $3.25$ &  $2.78 \pm 0.15$  &  $2.44 \pm 0.01$ & & & $2.75$ &  $2.41 \pm 0.10$ &  $2.15 \pm 0.09$ \\
$\thetapath$ & & & $0$ & $0$ &  $0.00$ & & & $>0$ & $>0$ & $1.07 \pm 0.08$ & & & $>0$ & $>0$ & $0.53 \pm 0.03$ \\
$\tpath$ & & & $2.00$ & $2.04 \pm 0.04$ & $2.00$ & & & $4.50$ & $3.85 \pm 0.30$ & $3.8 \pm 0.1$ & & & $3.50$ & $3.18 \pm 0.16$ & $3.1 \pm 0.1$  \\
$\thetatree$ & & & $0.00$ & $-0.04 \pm 0.04$ & $-0.10 \pm 0.2$ & & & $-1.14$ & $-0.92 \pm 0.17$ & $-0.96 \pm 0.02$ & & & $-0.80$ & $-0.53 \pm 0.02$ & $-0.57 \pm 0.08$ \\
$\ttree$ & & & $1.33$ & $1.33 \pm 0.04$ & $1.26 \pm 0.3$ & & & $2.17$ &  $1.92 \pm 0.15$ & $2.19 \pm 0.02$ & & & $1.83$ & $1.68 \pm 0.01$ & $1.75 \pm 0.01$ \\
\hline
\hline
\end{tabular}
\footnotetext[2]{Results for these ensembles were discussed in Ref.~\cite{RosaEveraersPRE2017}. They are reshown here for the purpose of comparison.}
\label{tab:exponents_df}
\end{table*}

Similarly to the scaling exponents for the expectation values of polymer observables, asymptotic values of for the pairs of exponents $(\theta,t)$ of Eq.~\eqref{eq:RdC-q} were also obtained through extrapolation to the large-tree limit.
More precisely, we followed closely the procedure described in Ref.~\cite{RosaEveraersPRE2017} which combines together the two extrapolation schemes:
\begin{enumerate}
\item
A fit of the data for $\thetal$ and $\thetatree$ 
to the following 3-parameter fit functions:
\begin{eqnarray}\label{eq:ExtrapolateThetalFuncts}
\log \thetal & = &     a + b N^{-\Delta_0} - b  (\Delta-\Delta_0) N^{-\Delta_0 } \log N \\
             & \equiv & a + b e^{-\Delta_0 \log N} - b  (\Delta-\Delta_0) e^{-\Delta_0 \log N} \log N
\end{eqnarray}
and
\begin{eqnarray}\label{eq:ExtrapolateThetaTreeFuncts}
\thetatree & =  &    a + b N^{-\Delta_0} - b  (\Delta-\Delta_0) N^{-\Delta_0 } \log N \\
           & \equiv & a + b e^{-\Delta_0 \log N} - b  (\Delta-\Delta_0) e^{-\Delta_0 \log N} \log N \\ 
\end{eqnarray}
and analogous expressions for $\tl$ and $\ttree$.
Eqs.~\eqref{eq:ExtrapolateThetalFuncts} and~\eqref{eq:ExtrapolateThetaTreeFuncts} correspond to a self-consistent linearisation of the 3 parameter fit
$\theta_{\ell,\text{tree}} = a + b \frac{1}{N^{\Delta}}$ around $\Delta = \Delta_0$.
We have carried out a one-dimensional search for
the value of $\Delta_0$ for which the fits yield vanishing $N^{-\Delta_0} \log N$ term.
Note that we have analyzed data for $\thetal$ (and $\tl$) in Eq.~\eqref{eq:ExtrapolateThetalFuncts} in the form $\log \thetal$ \textit{vs.} $\log N$ (log-log),
while for $\thetatree$ (and $\ttree$) we have used in Eq.~\eqref{eq:ExtrapolateThetaTreeFuncts} data in a log-linear representation, $\thetatree$ \textit{vs.} $\log N$. 
These two different functional forms have been found to produce the best (statistical significant) fits. 
\item
In the second method we fixed $\Delta =1$,
and we calculated the corresponding $2$-parameter best fits to the same data. 
\end{enumerate}
Results from the two fit procedures (including details such as the range of $N$'s considered for and the statistical significance of the fits) are summarized in Tables~\ref{tab:data-df_pl-l} and~\ref{tab:data-df_pr-r} in the Appendix,
as well as their averages (highlighted in boldface) which give our final estimates of scaling exponents. 
Unfortunately, for $\thetapath$ and $\tpath$ an analogous scheme can not be applied due to the limited ranges of $\ell$ available. 
In this last cases, our best estimates correspond to simple averages of single values (see boldfaced numbers in Table~\ref{tab:fit-df_prl-r}).
A summary of the exponents and final errors (calculated as in Sec.~\ref{sec:EstimatingScalExpsAvProps}) is given in Table~\ref{tab:exponents_df}.

\section{\label{sec:results}Results and discussion}

\subsection{\label{sec:NumericalResults}Average properties}

\subsubsection{Branching statistics}

%
\begin{figure}[htb!]
  \includegraphics[width=0.8\linewidth]{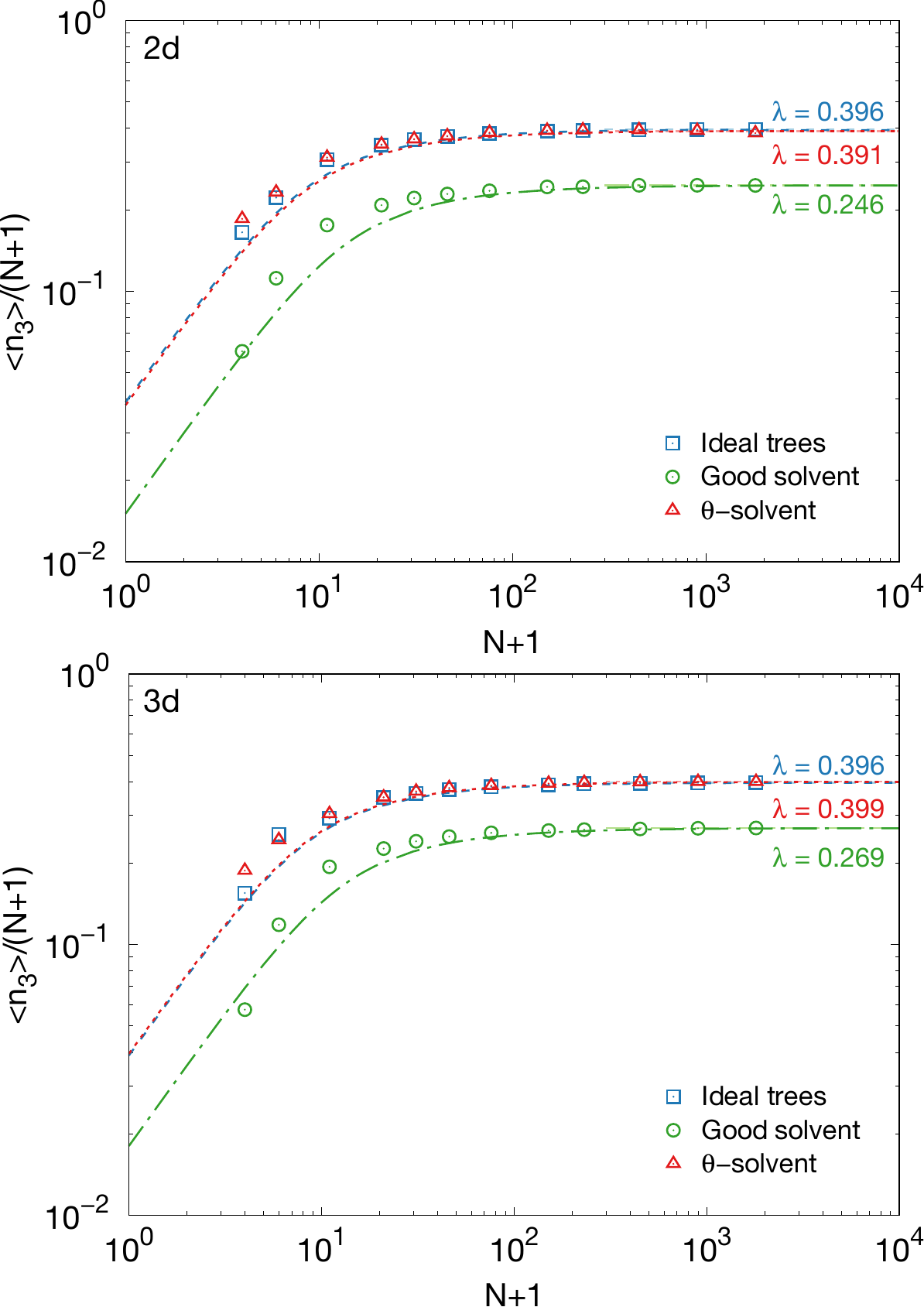}
  \caption{
  Average fraction of 3-functional nodes, $\langle n_3(N) \rangle / (N+1)$, as a function of the total number of tree nodes, $N+1$.
  Dashed lines correspond to the analytical expression for ideal trees by Daoud and Joanny, Eq.~(\ref{eq:DaoudJoanny}), with corresponding asymptotic branching probabilities $\lambda$.
  }
  \label{fig:n3-N}
\end{figure}

For every polymer ensemble considered in this work, we have computed the average number of branch points or 3-functional nodes of the tree structure, $\langle n_3(N)\rangle$, as a function of polymer size, $N$.
Numerical results for each $N$ are summarized in Table~\ref{tab:data-scaling_rho-epsilon} and plotted in Fig.~\ref{fig:n3-N}.

According to Daoud and Joanny~\cite{DaoudJoanny1981}, for ideal trees the ratio:
\begin{equation}\label{eq:DaoudJoanny}
  \frac{\langle n_3(N)\rangle}{\lambda N} \simeq
  \left\{
  \begin{array}{cc}
  1 \, , & \lambda N \gg 1 \\
  \lambda N \, , & \lambda N \ll 1
  \end{array}
  \right. ,
\end{equation}
where $\lambda$ defines the asymptotic branching probability per node. 
As first reported in Ref.~\cite{RosaEveraersJPhysA2016}, the formula by Daoud and Joanny summarizes well our data for ideal trees (see Fig.~\ref{fig:n3-N}) with branching probability $\lambda \approx 0.4$.
Interestingly, it describes well also the frequency of branching points in interacting trees,
either
self-avoiding trees in good solvent with $\lambda\approx0.246$ (in $2d$) and $\lambda\approx0.269$ (in $3d$)
or
trees in $\uptheta$-solvent. In particular, these last ones show again $\lambda \approx 0.4$ {\it i.e.} almost identical to the value of ideal polymers.
In spite of evident differences in their spatial structures (see Fig.~\ref{fig:structures}), we conclude that average branching in $\uptheta$-trees is mildly affected by volume interactions with respect to their ideal counterparts.

\subsubsection{Path length statistics}

%
\begin{figure*}[htb!]
  \includegraphics[width=0.7\linewidth]{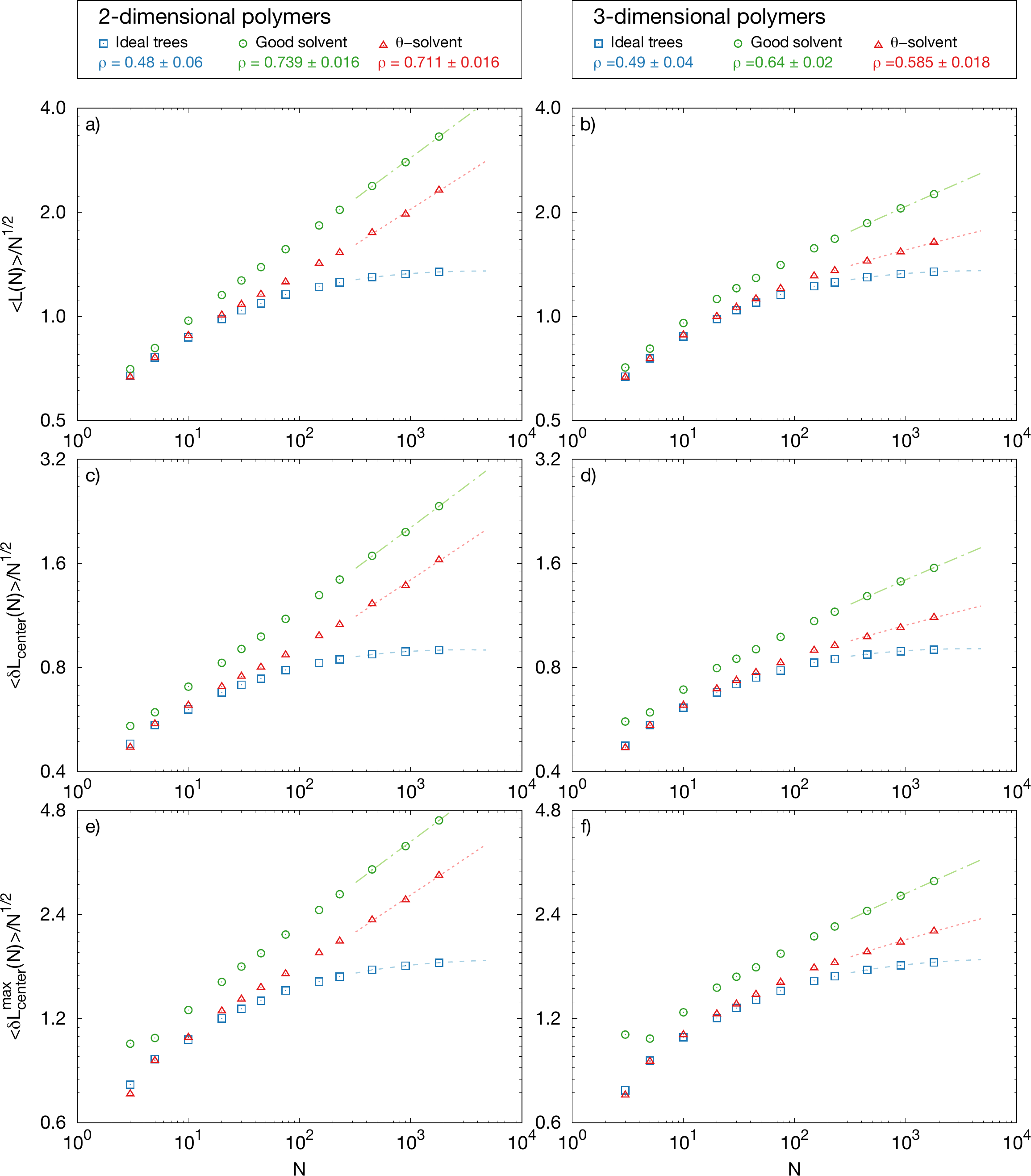}
  \caption{
  (a, b)
  Mean path distance between pairs of nodes, $\avLN$;
  (c, d) mean path distance between each node and the tree central node, $\braket{\dlcenter(N)}$, and
  (e, f) mean {\it longest} path distance of nodes from the tree central node, $\braket{\dlcentermax(N)}$.
  Left-hand-size plots correspond to $2d$ polymers, while r.h.s. plots are for $3d$ polymers.
  Straight dashed line correspond to the large-$N$ behavior $\avLN \sim \braket{\dlcenter(N)} \sim \braket{\dlcentermax(N)} \sim N^{\rho}$ with scaling exponent $\rho$.
  Data for $3d$ ideal polymers and $3d$ self-avoiding polymers in good solvent were already discussed in Ref.~\cite{RosaEveraersJPhysA2016} and shown here for comparison.
  }
  \label{fig:L-dlcenter-dlcentermax-N}
\end{figure*}

Fig.~\ref{fig:L-dlcenter-dlcentermax-N} and Table~\ref{tab:data-scaling_rho-epsilon} summarize the results for the different observables (see Table~\ref{tab:ListOfObservables}) introduced to quantify how path distances scale in term of the total tree weight $N$: 
(a) $\avLN$: the mean path distance between pairs of nodes;
(b) $\braket{\dlcenter(N)}$: the mean path distance between each node and the tree central node;
and
(c) $\braket{\dlcentermax(N)}$: the mean {\it longest} path distance of nodes from the tree central node.
These three observables are expected to scale with the tree polymer weight as $\sim N^\rho$, see Eq.~\eqref{eq:LvsN}.

Single estimated values for the scaling exponent $\rho$ were derived by applying the procedure described in Sec.~\ref{sec:EstimatingScalExpsAvProps} to each of these quantities,
with final results including error bars and the statistical significance being summarized in Table~\ref{tab:fit-scaling_rho-epsilon}.
Our final best estimates for $\rho$, obtained by averaging the separate results for $\avLN$, $\braket{\dlcenter(N)}$ and $\braket{\dlcentermax(N)}$, are given succinctly in Table~\ref{tab:exponents} and at the top of Fig.~\ref{fig:L-dlcenter-dlcentermax-N}, while in Table~\ref{tab:fit-scaling_rho-epsilon} they are highlighted in boldface with separate annotations for statistical and systematic errors.

\subsubsection{Weights of branches vs. path lengths}\label{sec:Nbr-dlroot-Ncenter-dlcenter}

%
\begin{figure*}[htb!]
  \centering
  \includegraphics[width=0.7\linewidth]{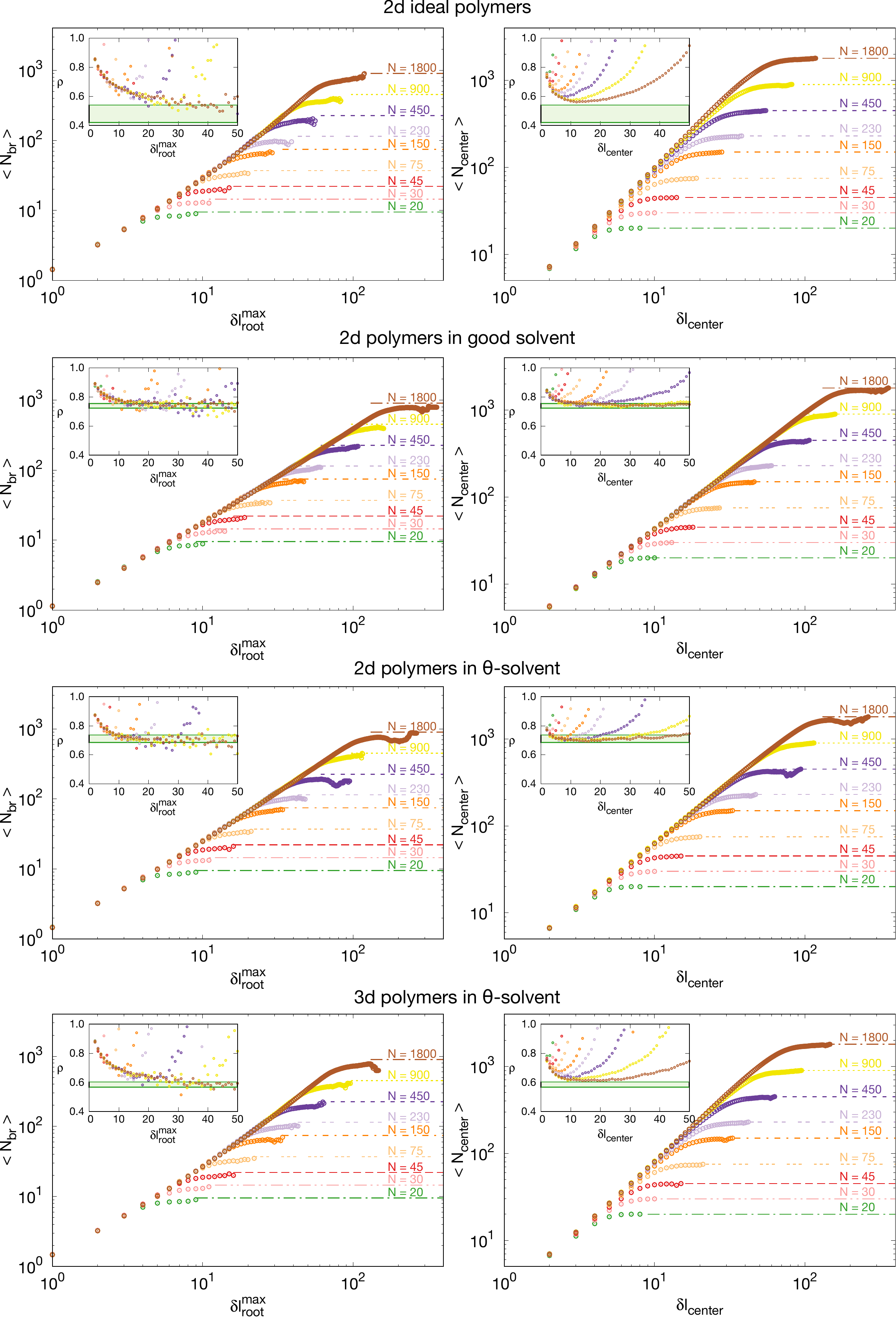}
  \caption{
  (Left) Average branch weight as a function of the longest contour distance to the branch root, $\braket{\Nbr(\dlmaxroot)}$.
  For large $\dlmaxroot$, curves saturate to the corresponding maximal branch weight $(N-1)/2$ (dashed horizontal lines). 
  (Right) Average branch weight, $\braket{\Ncenter(\dlcenter)}$, of paths whose distance from the central node does not exceed $\dlcenter$.
  For large $\dlcenter$, curves saturate to the corresponding total tree weight, $N$ (dashed horizontal lines).
  }
  \label{fig:Nbr-dlroot-Ncenter-dlcenter}
\end{figure*}

The scaling exponent $\rho$ describes also the functional relation between the average weight of branches of the trees and the typical path length of the branches.

In order to show this, we analyze
the behavior of the average branch weight as a function of the longest path length to the branch root, $\braket{\Nbr(\dlmaxroot)}$,
and
the average branch weight inside a given contour distance from the polymer central node, $\braket{\Ncenter(\dlcenter)}$.
In the limit of large trees ({\it i.e}, neglecting finite-chain effects) they should grow as:
\begin{eqnarray}
   \braket{\Nbr(\dlmaxroot)} & = & (\dlmaxroot)^{1/\rho} \, , \label{eq:Nbr-dlroot} \\
   \braket{\Ncenter(\dlcenter)} & = & (\dlcenter)^{1/\rho} \, . \label{eq:Ncenter-dlcenter}
\end{eqnarray}

We have computed $\braket{\Nbr(\dlmaxroot)}$ and $\braket{\Ncenter(\dlcenter)}$ for the different systems studied in this work, see Fig.~\ref{fig:Nbr-dlroot-Ncenter-dlcenter}.
As expected, in the large-$\dlmaxroot$ and large-$\dlcenter$ limits they plateau respectively to
$\braket{\Nbr(\dlmaxroot)} \rightarrow (N-1)/2$
and
$\braket{\Ncenter(\dlcenter)} \rightarrow N$.
At low and intermediate regimes, the scaling laws Eqs.~(\ref{eq:Nbr-dlroot}) and~(\ref{eq:Ncenter-dlcenter}) suggest that one can define a length dependent exponent $\rho = \rho(\dlmaxroot)$ as:
\begin{equation}\label{eq:LengthDependentRho}
  \rho(\dlmaxroot) \equiv \left[ \frac{\log\braket{\Nbr(\dlmaxroot+1)}-\log\braket{\Nbr(\dlmaxroot)}}{\log(\dlmaxroot+1)-\log(\dlmaxroot)} \right]^{-1} \, ,
\end{equation}
and an analogous expression for $\rho(\dlcenter)$.
The resulting values of the exponents are shown in the different insets of Fig.~\ref{fig:Nbr-dlroot-Ncenter-dlcenter}, they agree well within confidence intervals (shaded green area) with the corresponding final estimates for $\rho$ (see Table~\ref{tab:exponents}).

\subsubsection{Branch weight statistics}

%
\begin{figure}[htb!]
  \includegraphics[width=0.8\linewidth]{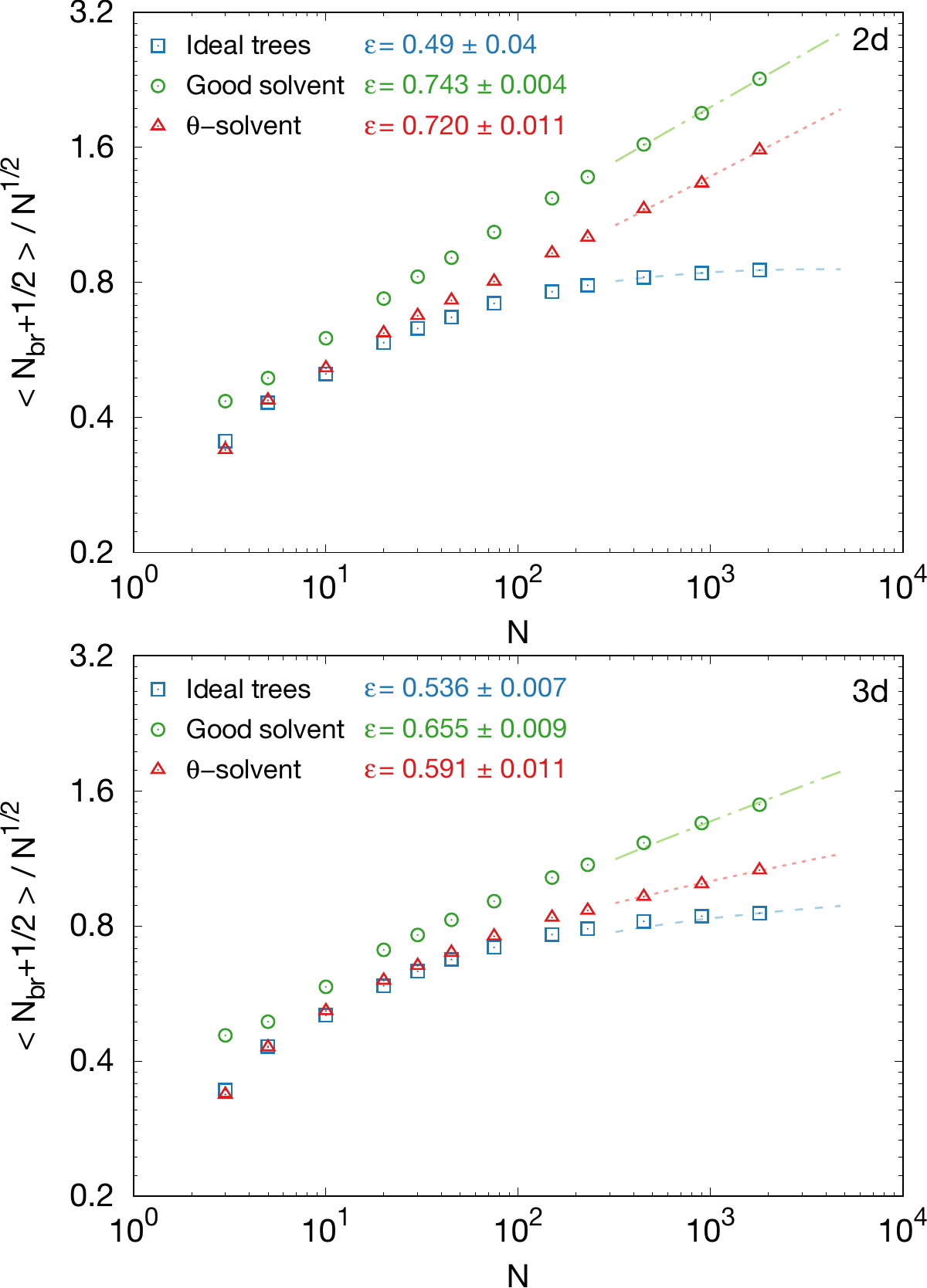}
  \caption{Average branch weight, $\braket{\Nbr(N)}$ as a function of the total tree mass, $N$.
  Straight dashed line correspond to the large-$N$ behavior $\braket{\Nbr(N)} \sim N^{\epsilon}$ with scaling exponents $\epsilon$.
  Data for $3d$ ideal polymers and $3d$ self-avoiding polymers in good solvent were discussed in previous work~\cite{RosaEveraersJPhysA2016}.
  }
  \label{fig:Nbr-N}
\end{figure}

As anticipated by Eq.~\eqref{eq:NbrvsN}, the average weight of the polymer branches $\braket{\Nbr}$ scales with the size $N$ in terms of the characteristic exponent $\epsilon$.
The asymptotic behaviors for each tree ensemble are reported in Fig.~\ref{fig:Nbr-N}, while detailed values for each $N$ are summarized in Table~\ref{tab:data-scaling_rho-epsilon}. 
Final estimates for the exponent $\epsilon$ are indicated both in the figure and in Table~\ref{tab:exponents}, while Table \ref{tab:fit-scaling_rho-epsilon} summarizes the statistical details of the fits.
We stress that the mathematical relation $\rho=\epsilon$ first pointed out by Janse van Rensburg and Madras~\cite{JansevanRensburgMadrasJPhysA1992} for trees in good solvent is accurately verified by all systems studied in this work, and in particular also for $\uptheta$-trees.

\subsubsection{Conformational statistics of linear paths}\label{sec:scaling_nupath}

%
\begin{figure*}[htb!]
  \includegraphics[width=0.75\linewidth]{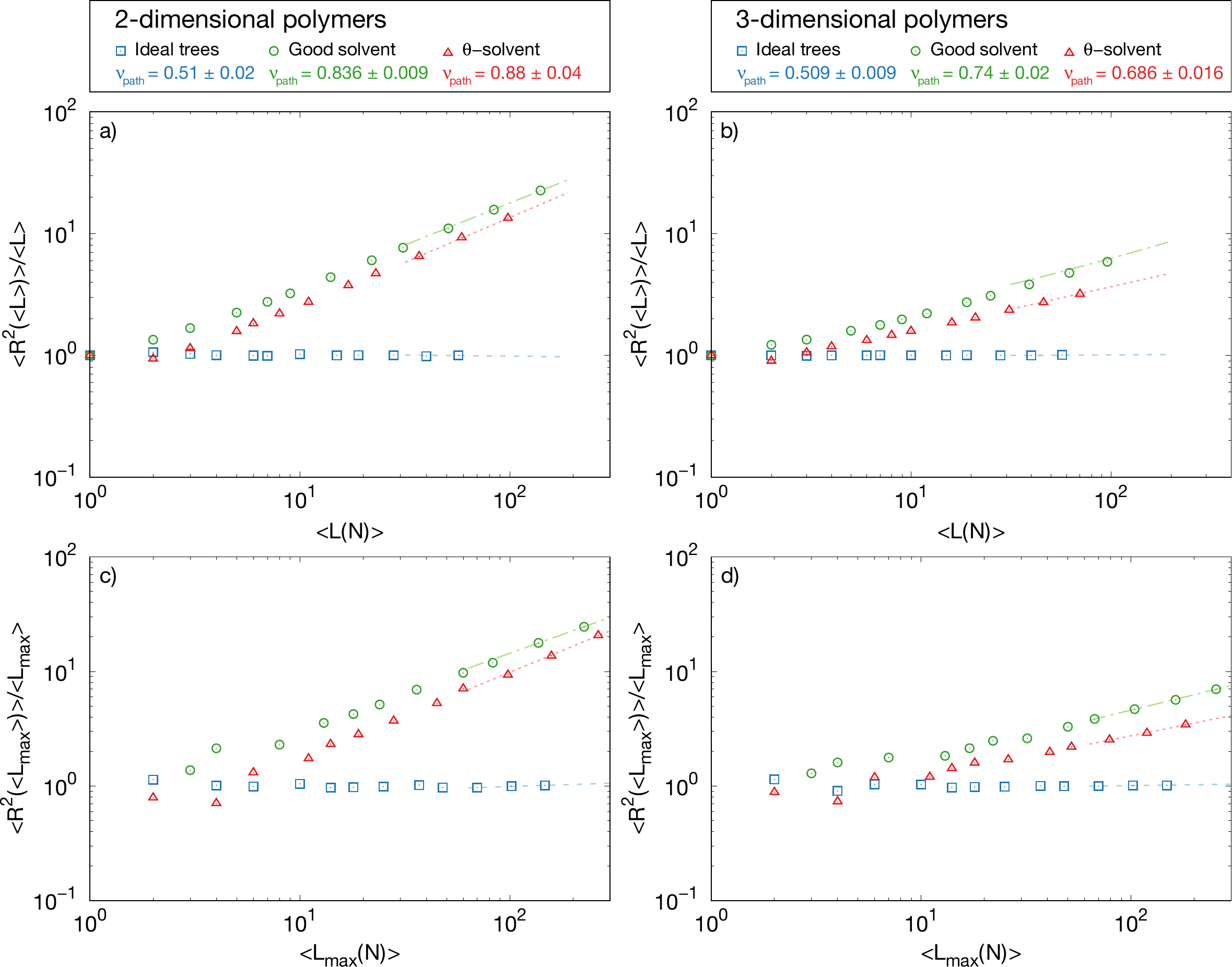}
  \caption{
  (a,b) Mean square end-to-end spatial distance, $\avRl$, of paths of length $\ell=\braket{L(N)}$;
  (c, d) mean square end-to-end spatial distance, $\braket{R^{2}(\Lmax)}$, of the longest paths.
  L.h.s. plots correspond to $2d$ polymers, while r.h.s. plots are for $3d$ polymers.
  Straight dashed lines correspond to the large-$N$ scaling behavior, Eqs.~(\ref{eq:R2-L}) and~(\ref{eq:R2-Lmax}), with exponent $\nupath$.
  Data for $3d$ ideal polymers and $3d$ self-avoiding polymers in good solvent were discussed in previous work~\cite{RosaEveraersJPhysA2016}.
  }
  \label{fig:R2-L-Lmax}
\end{figure*}

To study the conformational statistics of linear tree paths and determine the corresponding scaling exponent $\nu_{\rm path}$ (see Eq.~(\ref{eq:R2vsL})),
we have analyzed the end-to-end mean-square spatial distance of paths of average length $\langle L (N) \rangle$, and of average maximal length, $\braket{\Lmax(N)}$.
For the scaling relationship Eq.~\eqref{eq:R2vsL}, we expect that corresponding mean-square end-to-end spatial distances of these paths obey:
\begin{align}
   \braket{R^{2}(\braket{L(N)})} & \sim \avLN^{2\nupath} \, , \label{eq:R2-L} \\
   \braket{R^{2}(\braket{\Lmax(N)})} & \sim \braket{\Lmax(N)}^{2\nupath} \, . \label{eq:R2-Lmax}
\end{align}
Detailed results for our systems are summarized in Table~\ref{tab:data-scaling_nu-nupath} and shown in Fig.~\ref{fig:R2-L-Lmax}, together with the final estimates for the exponent $\nupath$ (see Table~\ref{tab:exponents}).
The complete statistical details about the fits used to calculate the exponent from the data are summarized in Table~\ref{tab:fit-scaling_nu-nupath}.

\begin{figure*}[htb!]
  \centering
  \includegraphics[width=0.75\linewidth]{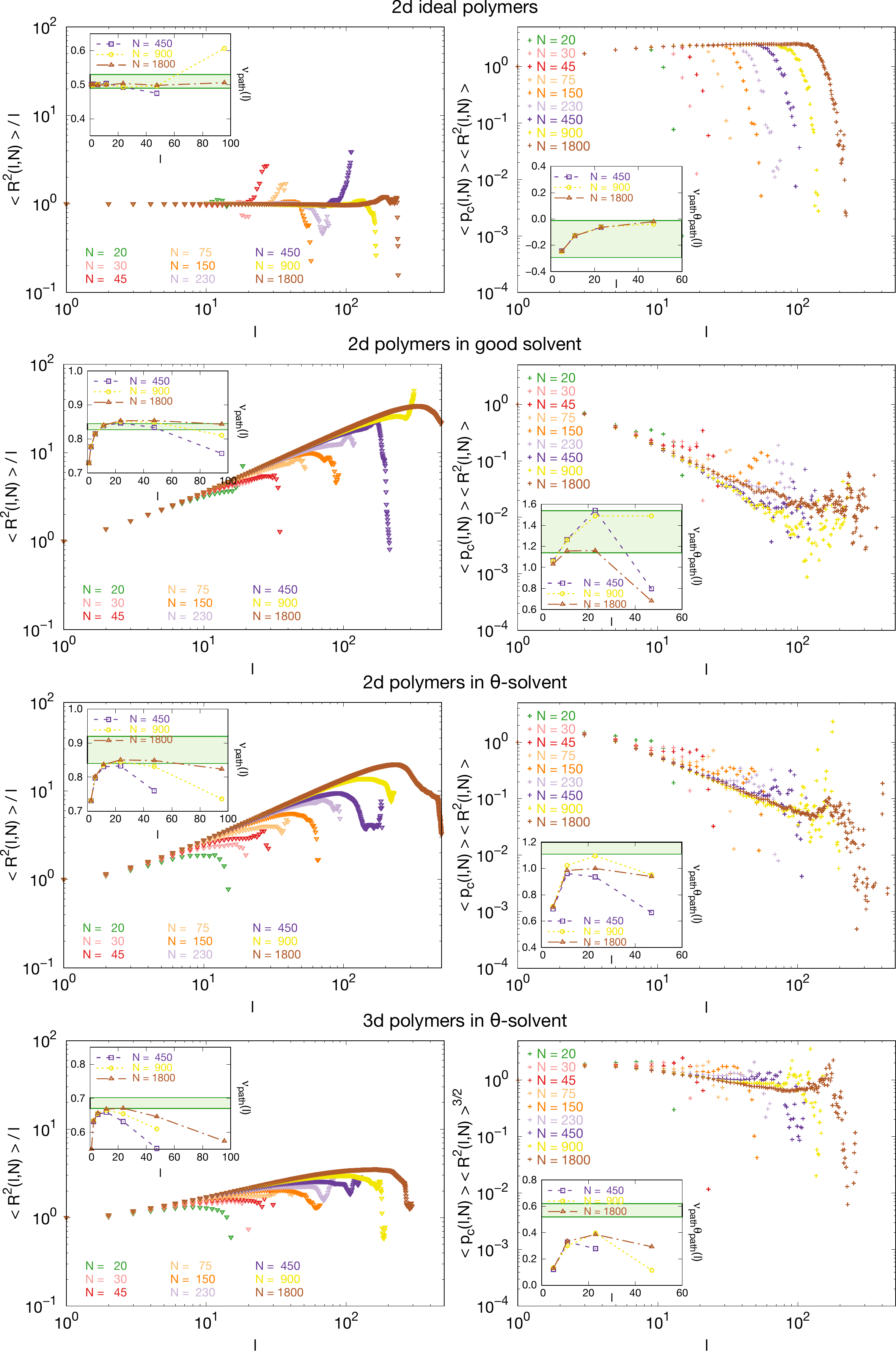}
  \caption{
  (Left) Mean-square end-to-end distance, $\avRlN$, of linear paths of length $\ell$.
  Insets: scaling exponent $\braket{\nupath(\ell)}$, averaged over log-spaced intervals for polymer size $N \geq 450$.
  (Right) Mean closure probabilities, $\avpclN$, between ends of linear paths of length $\ell$ normalized to the mean-field expectation value $\avRlN^{d/2}$.
  Insets: scaling exponent $\braket{\nupath\thetapath(\ell)}$, averaged over log-spaced intervals for polymer size $N \geq 450$.
    }
  \label{fig:R2-pc-l}
\end{figure*}

Then, we have measured the mean-square end-to-end distances of linear paths of length $\ell$ for trees of weight $N$, $\avRlN$, see the l.h.s panels in Fig.~\ref{fig:R2-pc-l}.
As for quantities $\braket{\Nbr(\dlmaxroot)}$ and $\braket{\Ncenter(\dlcenter)}$ (see Eq.~(\ref{eq:LengthDependentRho})), 
we define a length-dependent scaling exponent $\nupath(\ell)$ through the numerical slope in log-log scale: 
\begin{equation}
  \nupath(\ell) \equiv \frac{1}{2}\frac{\log\braket{R^2(\ell+1)}-\log\braket{R^2(\ell)}}{\log(\ell+1)-\log(\ell)} \, .
\end{equation}
Insets in the l.h.s. plots in Fig.~\ref{fig:R2-pc-l} show the values of $\nupath = \nupath(\ell)$ for the longest polymer sizes $N\geq450$, after having been averaged over log-spaced intervals of $\ell$ for enhancing the visualization. 
The green shaded region indicates the confidence intervals of the final estimates of $\nupath$ summarized in Table~\ref{tab:exponents}.

Finally, we conclude this analysis on the spatial conformations of linear paths by discussing the scaling for the average closure probabily, $\avpclN$, namely the average number of contacts between pairs of nodes separated by a contour distance $\ell$ within the tree polymer structure of weight $N$.
Mean-field considerations~\cite{RosaEveraersJPhysA2016,RosaEveraersJCP2016} suggest that $\avpcl$ should scale as $\sim \braket{R^2(\ell)}^{-d/2} \sim \ell^{-d\nupath}$.
Not surprisingly, r.h.s. plots in Fig.~\ref{fig:R2-pc-l} show that only ideal polymers obey the mean-field result while interacting polymers display significant deviations which, as anticipated by Eq.~(\ref{eq:PCvsL}), can be quantified by introducing the additional scaling exponent $\thetapath$, 
$\avpcl \sim \ell^{-\nupath(d+\thetapath)}$.
Again, we calculated the slope of the data at low and intermediate values of $\ell$ for trees with $N\geq450$ and averaged the results over log-spaced intervals to compute the product $\nupath\thetapath (\ell)$.
Then, we took the results for $N=1800$ and the values of $\nupath$ derived previously from Eqs.~(\ref{eq:R2-L}) and~(\ref{eq:R2-Lmax}) in order to best estimate the corresponding values for $\thetapath$.
Final results for the different tree ensembles are summarized in Table~\ref{tab:exponents}.

\subsubsection{Conformational statistics of lattice trees}\label{sec:ConfStatTrees}

%
\begin{figure}[htb!]
  \includegraphics[width=0.75\linewidth]{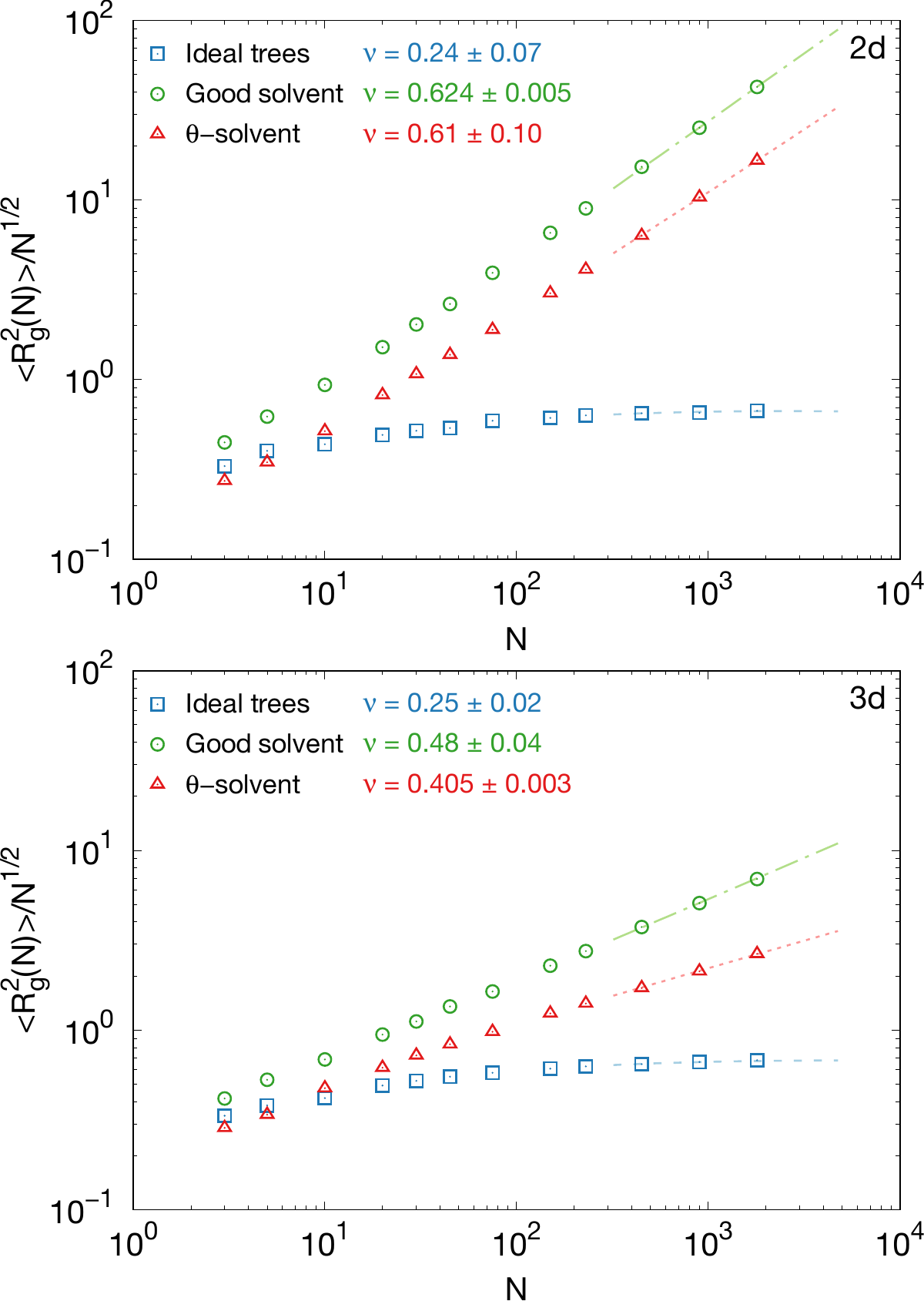}
  \caption{
  Mean square gyration radius, $\avRgN$.
  Straight lines correspond to the large-$N$ behavior $\avRgN \sim N^{2\nu}$ with scaling exponent $\nu$.
  Data for $3d$ ideal polymers and $3d$ self-avoiding polymers in good solvent were discussed in previous work~\cite{RosaEveraersJPhysA2016}.
  }
  \label{fig:Rg2-N}
\end{figure}
\begin{figure}[htb!]
  \includegraphics[width=0.75\linewidth]{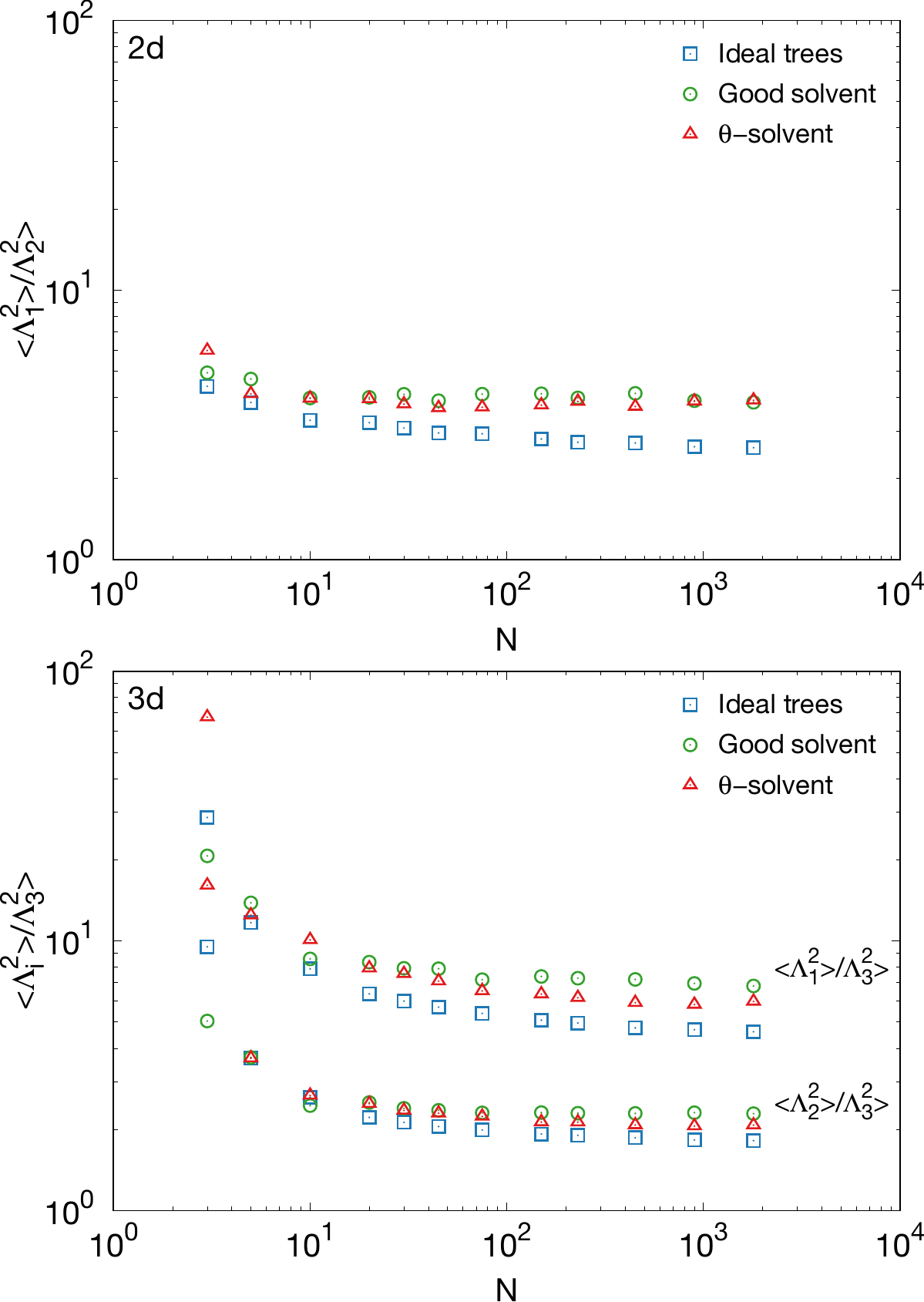}
  \caption{
  Average polymer aspect ratios from the expectation values of the gyration tensor eigenvalues, $\braket{\Lambda_{1}^{2}} > \braket{\Lambda_{2}^{2}} > \braket{\Lambda_{3}^{2}}$.
  Data for $3d$ ideal polymers and $3d$ self-avoiding polymers in good solvent were discussed in previous work~\cite{RosaEveraersJPhysA2016}.
  }
  \label{fig:lambda-N}
\end{figure}

Trees spatial conformations were analyzed in terms of the scaling behavior of the expectation value of the square gyration radius with the total tree weight $N$ (see Eq.~(\ref{eq:Rg2vsN})):
\begin{equation}\label{eq:DefineRg2vsN}
\avRgN \equiv \left\langle \frac{1}{N+1} \sum_{i=1}^{N+1} (\vec r_i - \vec r_{\rm cm})^2 \right \rangle \sim N^{2\nu} \, ,
\end{equation}
where $\vec r_i$ is the spatial position of the $i$-th monomer of the tree and $\vec r_{\rm cm} \equiv 1/(N+1) \sum_{i=1}^{N+1} \vec r_i$ is the spatial position of the tree centre of mass.
$\avRgN$ quantifies polymer swelling in the presence of the solvent~\cite{RubinsteinColbyBook} with respect to the ideal conditions. 

Numerical results for the different ensembles are summarized in Table~\ref{tab:data-scaling_nu-nupath} and shown in Fig.~\ref{fig:Rg2-N}. 
We notice, in particular, that good solvent conditions exhibit the largest swelling while $\uptheta$-polymers exhibit intermediate behavior between ideal and good solvent conditions due to the balance between repulsive and attractive monomer-monomer spatial interactions, Eq.~(\ref{eq:Hint}). 
This observation agrees with the corresponding estimates for the scaling exponent $\nu$ whose values and confidence intervals are shown in Fig.~\ref{fig:Rg2-N} and Table~\ref{tab:exponents},
while the numerical details about their derivation based on best fits to the data are summarized in Table~\ref{tab:fit-scaling_nu-nupath}.
The result for $\uptheta$-polymers in $2d$ is, in particular, in good agreement with the accurate value $\nu = 0.5359\pm0.0003$ measured by Hsu and Grassberger~\cite{HsuGrassbergerJStatMech2005}.
It is also worth pointing out that polymer swelling from ideal to $\uptheta$-solvent conditions as the result of monomer-monomer interactions is mainly affecting the average polymer size while, in comparison, the average internal connectivity appears virtually unperturbed, see Fig.~\ref{fig:n3-N}.

Finally, we have considered the average polymer shape which can be quantified in terms of the expectation values of the ordered eigenvalues,
$\braket{\Lambda_{1}^{2}(N)} > \braket{\Lambda_{2}^{2}(N)} > \braket{\Lambda_{3}^{2}(N)} \rangle$ with $\sum_{i=1}^3 \braket{\Lambda_{i}^{2}(N)} = \avRgN$,
of the $3\times3$ (in $d=3$) symmetric polymer gyration tensor, $T=T(N)$, whose components are given by:
\begin{equation}\label{eq:DefineGyrTensor}
T_{\alpha\beta}(N) \equiv \frac{1}{N+1} \sum_{i=1}^{N+1} (\vec r_{i,\alpha} - \vec r_{{\rm cm},\alpha}) (\vec r_{i,\beta} - \vec r_{{\rm cm},\beta}) \, ,
\end{equation}
where $\vec r_{i,\alpha=x,y,z}$ and $\vec r_{{\rm cm},\beta=x,y,z}$ stand for the spatial components of the corresponding vectors.
Analogous expressions hold for polymers in $d=2$. 
Fig.~\ref{fig:lambda-N} shows the aspect ratios $\braket{\Lambda_{i}^{2}}/\braket{\Lambda_{3}^{2}}$ for $i=1,2$,
in particular we may notice that both $2d$ and $3d$ ideal polymers appear slightly less a-spherical than their interacting counterparts.

\subsection{Distribution functions}

As anticipated in Sec.~\ref{sec:TheoryDistributions},
we conclude this study on the statistical physics properties of lattice trees by discussing the distribution functions for trees connectivity and spatial conformations.
In particular, we show that these functions have universal shapes which can be described in terms of the Redner-des Cloizeaux (RdC) theory:
the latter implies the existence of new sets of exponents which can be quantitatively related to the exponents describing the scaling of average properties~(\ref{eq:Rg2vsN})-(\ref{eq:PCvsL}) through generalized Fisher-Pincus (FP) relationships.

\subsubsection{Path length statistics}

%
\begin{figure}[htb!]
  \includegraphics[width=\linewidth]{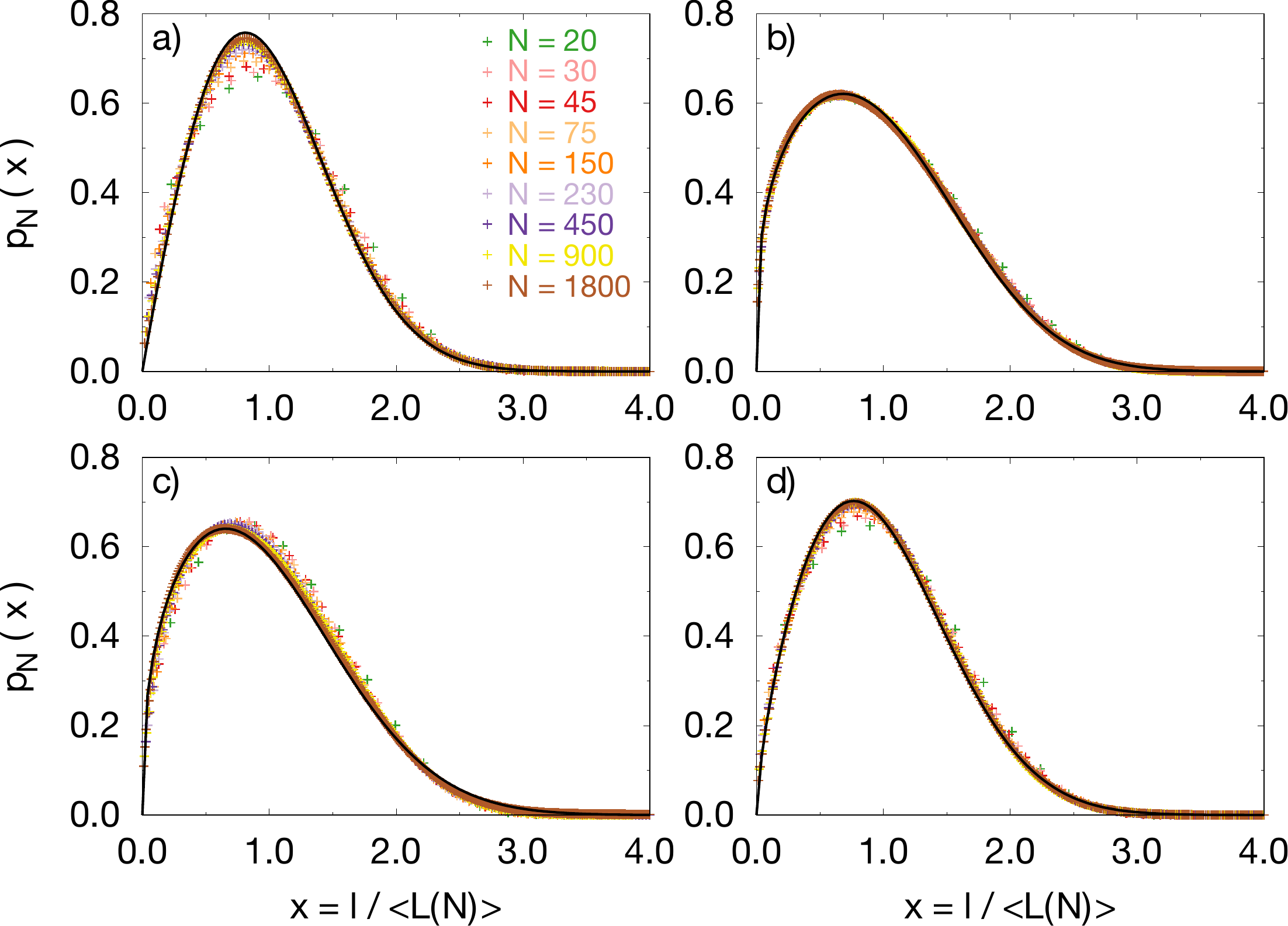}
  \caption{Distribution functions $\pl$ of linear paths of length $\ell$ for polymers of size $N$:
  (a) $2d$ ideal polymers,
  (b) $2d$ polymers in good solvent,
  (c) $2d$ polymers in $\uptheta$-solvent and
  (d) $3d$ polymers in $\uptheta$-solvent.
  Black solid lines correspond to the Redner-des Cloizeaux function Eq.~\eqref{eq:RdC-q} computed with the final parameters $(\thetal,\tl)$ summarized in Table \ref{tab:exponents_df}.
  }
  \label{fig:pl-l}
\end{figure}

We discuss first the distribution functions of linear paths of length $\ell$ for polymers of size $N$, $\pl$.
As shown in Fig.~\ref{fig:pl-l}, this function obeys the universal RdC scaling form 
described by Eq.~\eqref{eq:RdC-pl}.
We calculated then the pair of exponents $(\thetal, \tl)$ for each polymer weigth of $N$ by fitting the plotted data with rescaled path lengh $x\equiv\ell/\avLN$ to Eq.~\eqref{eq:RdC-q} (see Table~\ref{tab:data-df_pl-l} for details).
Then, we obtained the final estimation for the pair $(\thetal, \tl)$ by extrapolating the finite-$N$ results using Eq.~\eqref{eq:ExtrapolateThetalFuncts} and the related methods explained in Sec.~\ref{sec:EstimatingScalExpsDistrFuncts}:
the reconstructed distribution functions are shown for comparison as black curves in Fig.~\ref{fig:pl-l}.

Finally, we compared the extrapolated values for $(\thetal, \tl)$ to the quantitative Fisher-Pincus (FP) expressions (Eqs.~(\ref{eq:FP-thetal}) and~(\ref{eq:FP-tl})) relating those exponents to $\rho$. 
This task has been summarized in Table~\ref{tab:exponents_df} showing the final estimations for $(\thetal, \tl)$ and the results from the FP formulas with $\rho$ being equal to the Flory values and to the final values estimated from the present simulations. 

Although we may appreciate the power of the Flory theory which, once again, in spite of its simplicity seems to agree well with the data in a close-to-quantitative manner,
we may also notice that, in all non-ideal cases, it underestimates systematically the exponent $\thetal$ and overestimates $\tl$, with differences between theory and simulations more pronounced in $d=2$.
Finally, the exponents $(\thetal, \tl)$ agree well with the predictions of the FP relations with $\rho$ corresponding to the values estimated in this work.

\subsubsection{Conformational statistics of linear paths}

%
\begin{figure}[htb!]
  \includegraphics[width=\linewidth]{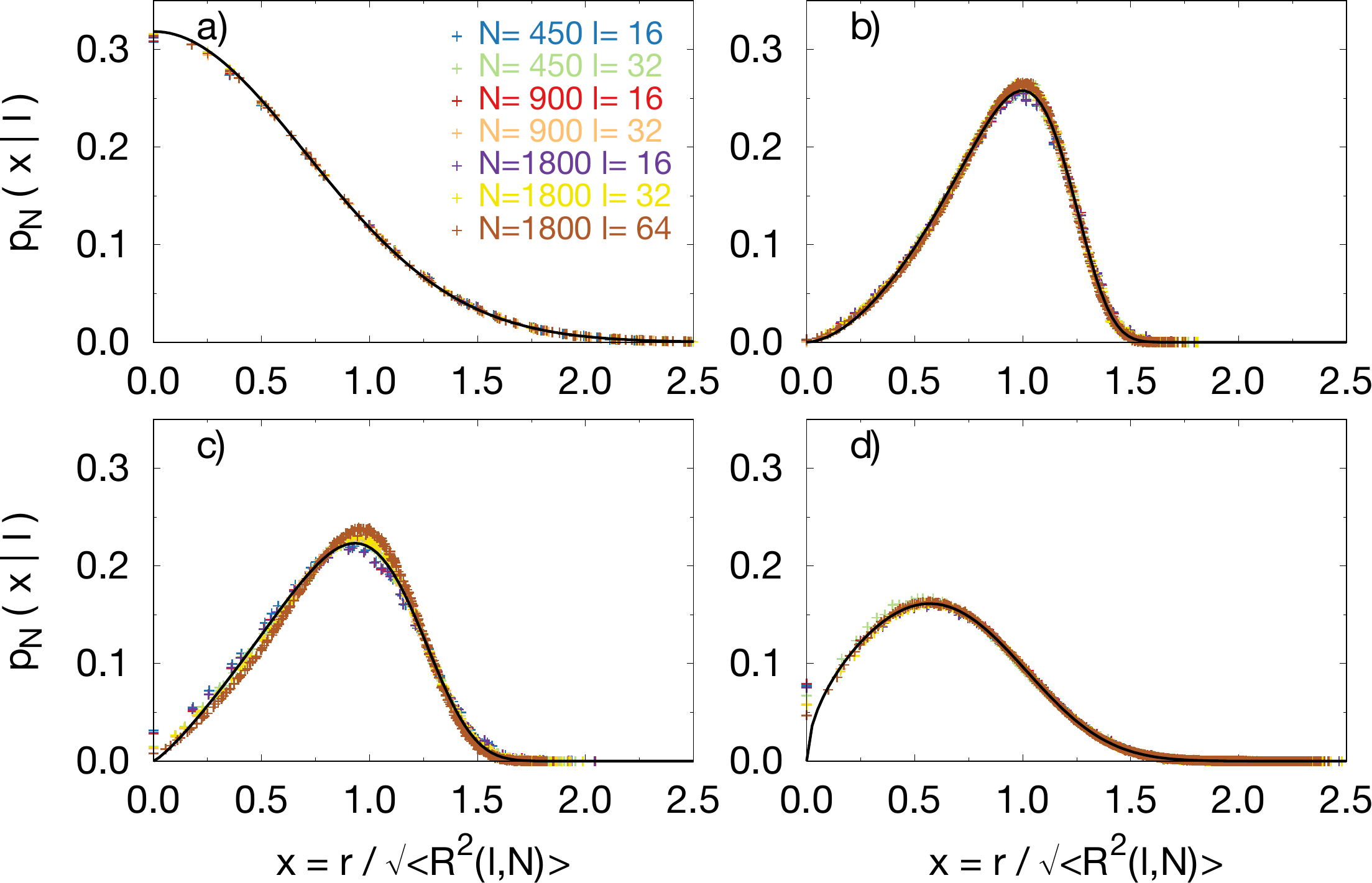}
  \caption{Distribution functions $\prl$ of end-to-end spatial vectors $\vec r$ for linear paths of length $\ell$ on polymers of size $N$:
  (a) $2d$ ideal polymers,
  (b) $2d$ polymers in good solvent,
  (c) $2d$ polymers in $\uptheta$-solvent and
  (d) $3d$ polymers in $\uptheta$-solvent.
  Black solid lines correspond to the Redner-des Cloizeaux function Eq.~\eqref{eq:RdC-q} computed by using the final estimations of parameters $(\thetapath,\tpath)$ summarized in Table~\ref{tab:exponents_df}.
  }
  \label{fig:prl-r}
\end{figure}

We considered the distribution function $p_N(\vec r|\ell)$ of end-to-end vectors of linear paths of lengh $\ell$ on trees of weight $N$ and we plot them as a function of the scaling variable $x \equiv |\vec{r}| / \sqrt{\avRlN}$.
As shown in Fig.~\ref{fig:prl-r} the data from different $N$'s collapse to a single universal shape,
which obeys the RdC functional form~\eqref{eq:RdC-q} with given exponents $\thetapath$ and $\tpath$.
In the ideal $2d$ case, $p_N(\vec r|\ell)$ is well described by the Gaussian function with $\thetapath=2$ and $\tpath=0$ (black curve in Fig.~\ref{fig:prl-r}(a)). 
In the other cases, the exponents $\thetapath$ and $\tpath$ have been found by best fits of the RdC expression to the data for different $N$'s and $\ell$'s (see Table~\ref{tab:fit-df_prl-r}).
As briefly mentioned in Sec.~\ref{sec:EstimatingScalExpsDistrFuncts}, our final estimations for the pairs $(\thetapath, \tpath)$ were given by averaging the single values determined for trees with $N\geq450$ and path lengths $\ell=16, 32, 64$, the corresponding RdC functions shown as black curves in panels (b)-(d) of Fig.~\ref{fig:prl-r}.
In fact, no extrapolation was attempted in this case due to the limited range of path lengths available for our systems. 

Finally, in Table~\ref{tab:exponents_df} we compare the estimated values for $\tpath$ to the Fisher-Pincus (FP) relationship from Eq.~(\ref{eq:FP-tpath}) upon substitution of theoretical Flory values and our numerical results.
The agreement is overall good, with values from the Flory theory typically overestimating the numerical predictions.
It was remarked~\cite{RosaEveraersPRE2017} that $\thetapath>0$, otherwise there seems to be no relation between this exponent and the others: 
this suggests
that $\thetapath$ should be regarded as a genuinely novel exponent.

\subsubsection{Conformational statistics of lattice trees}

%
\begin{figure}[htb!]
  \includegraphics[width=\linewidth]{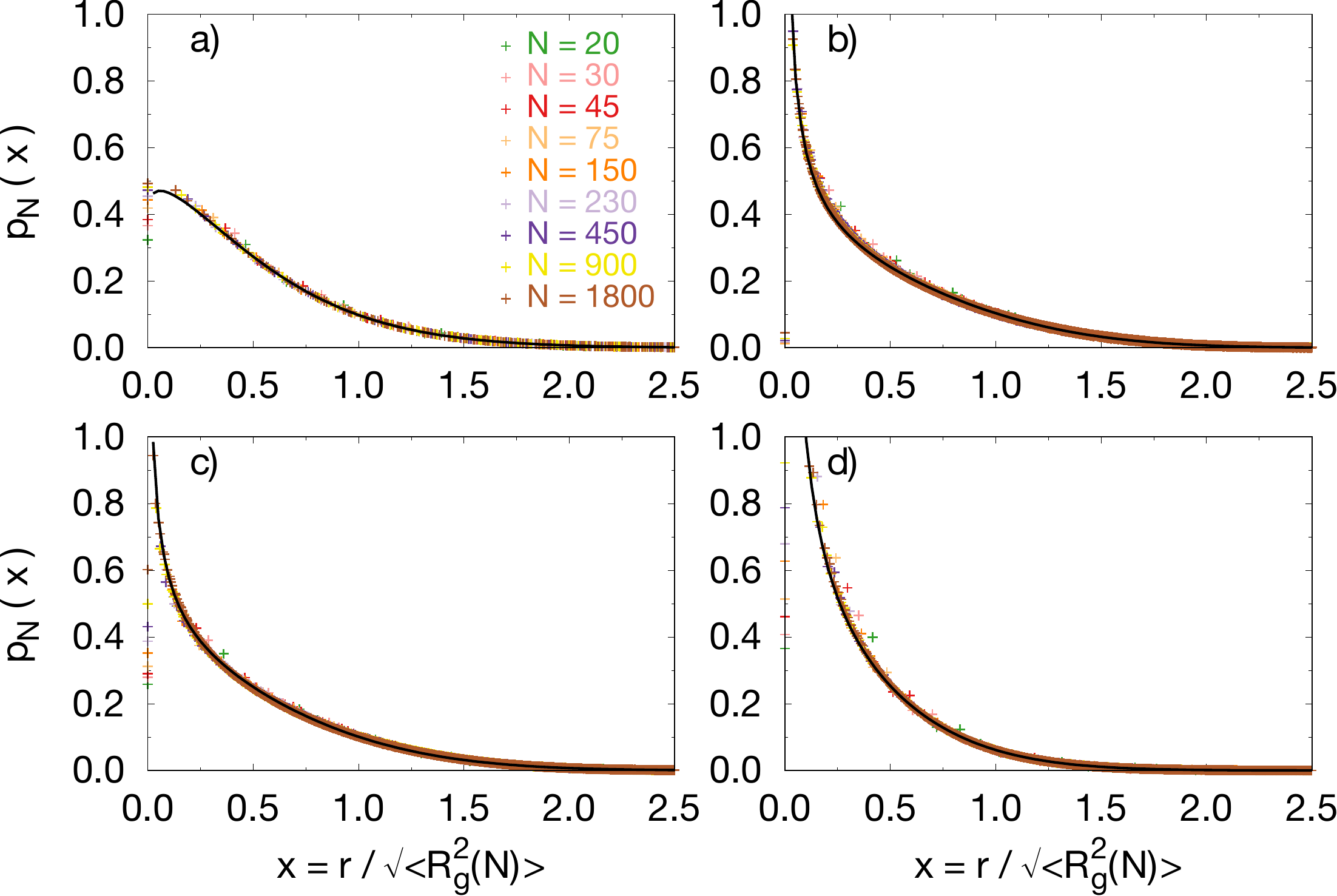}
  \caption{Distribution functions $\pr$ of end-to-end spatial vectors $\vec r$ between pairs of nodes on tree polymers of size $N$:
  (a) $2d$ ideal polymers,
  (b) $2d$ polymers in good solvent,
  (c) $2d$ polymers in $\uptheta$-solvent and
  (d) $3d$ polymers in $\uptheta$-solvent.
  Black solid lines correspond to the Redner-des Cloizeaux function Eq.~\eqref{eq:RdC-q} computed by using the final estimations of parameters $(\thetatree,\ttree)$ summarized in Table~\ref{tab:exponents_df}.
  }
  \label{fig:pr-r}
\end{figure}

Finally, we conclude this analysis by considering the distribution functions for the end-to-end distances between pairs of nodes of trees of weight $N$, $\pr$, as a function of the rescaled distance $x\equiv|\vec{r}|/\sqrt{2\avRgN}$.
As in the previous cases, the curves from different polymers superimpose (see Fig.~\ref{fig:pr-r}) and agree with the Redner-de Cloizeaux functional form Eq.~\eqref{eq:RdC-q}.
The corresponding pairs of exponents $(\thetatree,\ttree)$ were calculated by first fitting data to~\eqref{eq:RdC-q} for every $N$ (see Table~\ref{tab:data-df_pr-r}) and then extrapolating the results according to the numerical procedure outlined in Sec.~\ref{sec:EstimatingScalExpsDistrFuncts}.
Final results are shown in Table~\ref{tab:exponents_df} and compared to the predictions of Fisher-Pincus relations Eqs.~(\ref{eq:FP-thetatree}) and~(\ref{eq:FP-ttree}) upon substitutions of the theoretical (Flory) and numerical values for $\thetapath$ and $\nu$.

\section{Conclusions}\label{sec:conclusions}

Following the outline of recent work~\cite{RosaEveraersJPhysA2016,RosaEveraersJCP2016,RosaEveraersPRE2017} by our group,
in the present paper we have presented a systematic, quantitative analysis regarding the scaling properties of single conformations of branching polymers in $\uptheta$-solvent conditions and annealed connectivity.
To this purpose, we have employed a mix of theoretical considerations based on the Flory theory, results of on-lattice Monte Carlo computer simulations and rigorous numerical extrapolation methods to derive best estimates for scaling exponents of average chain properties and distribution functions.

We highlight three main crucial aspects of the present article.

First,
the need to generalize the Monte Carlo method first discussed in Ref.~\cite{RosaEveraersJPhysA2016,RosaEveraersJCP2016} in order to model properly the effects of the $\uptheta$-solvent.
This task (Sec.~\ref{sec:ThetaInteraction}) has been accomplished by suitably tailoring the force-field in order to recover the most accurate value to date~\cite{Madras1997} of the scaling exponent $\nu$ relating the average polymer size and the polymer weight $N$ in $3d$, see Eq.~(\ref{eq:Rg2vsN}). 
Remarkably, we found that the same interaction parameters found for the $3d$ case appear appropriate also for $2d$ polymers (Sec.~\ref{sec:ConfStatTrees}) without additional fine calibration: this shows the generality and robustness of our model. 

Second,
we performed a detailed analysis of the average properties, Eqs.~(\ref{eq:Rg2vsN})-(\ref{eq:PCvsL}), of tree polymers focusing, in particular, on the accurate derivation of the corresponding scaling exponents.
This analysis, interesting {\it per se}, has been carried out by systematically comparing our data to the predictions of the Flory theory.
The study, which confirms and completes the discussion first started in Refs.~\cite{RosaEveraersJPhysA2016,RosaEveraersJCP2016}, supports the idea that simple Flory theories~\cite{Isaacson1980,MaritanGiacometti2013,EveraersGrosbergRubinsteinRosaSoftMatter2017} can indeed be used to gain insight into the physics of polymers with branched architectures, in particular in rationalizing the trends of critical quantities with respect to the typical tree path length or polymer mass (see Table~\ref{tab:exponents}).

Third,
by measuring and discussing the complete statistics of path lengths and spatial polymer conformations we were able to move beyond the Flory approximation and, thus, fill the remaining gaps. 
In particular, we confirmed that the distribution functions obey the Redner-des Cloizeaux statistics (Eq.~\eqref{eq:RdC-q}) with novel exponents which can be quantitatively understood by suitably generalizing the classical Fisher-Pincus theory of polymer physics (see Table~\ref{tab:exponents_df}). 

Taken together, the results of our work draw a complete picture of the physics of polymers with annealed branching architectures in $\uptheta$-solvent conditions.

\begin{acknowledgments}
The authors acknowledge computational resources from SISSA HPC-facilities.

\end{acknowledgments}


%
\clearpage
\appendix*
\section{Supplemental Information}\label{sec:SI}
\begin{figure*}[htb!]
  \centering
  \includegraphics[width=0.75\linewidth]{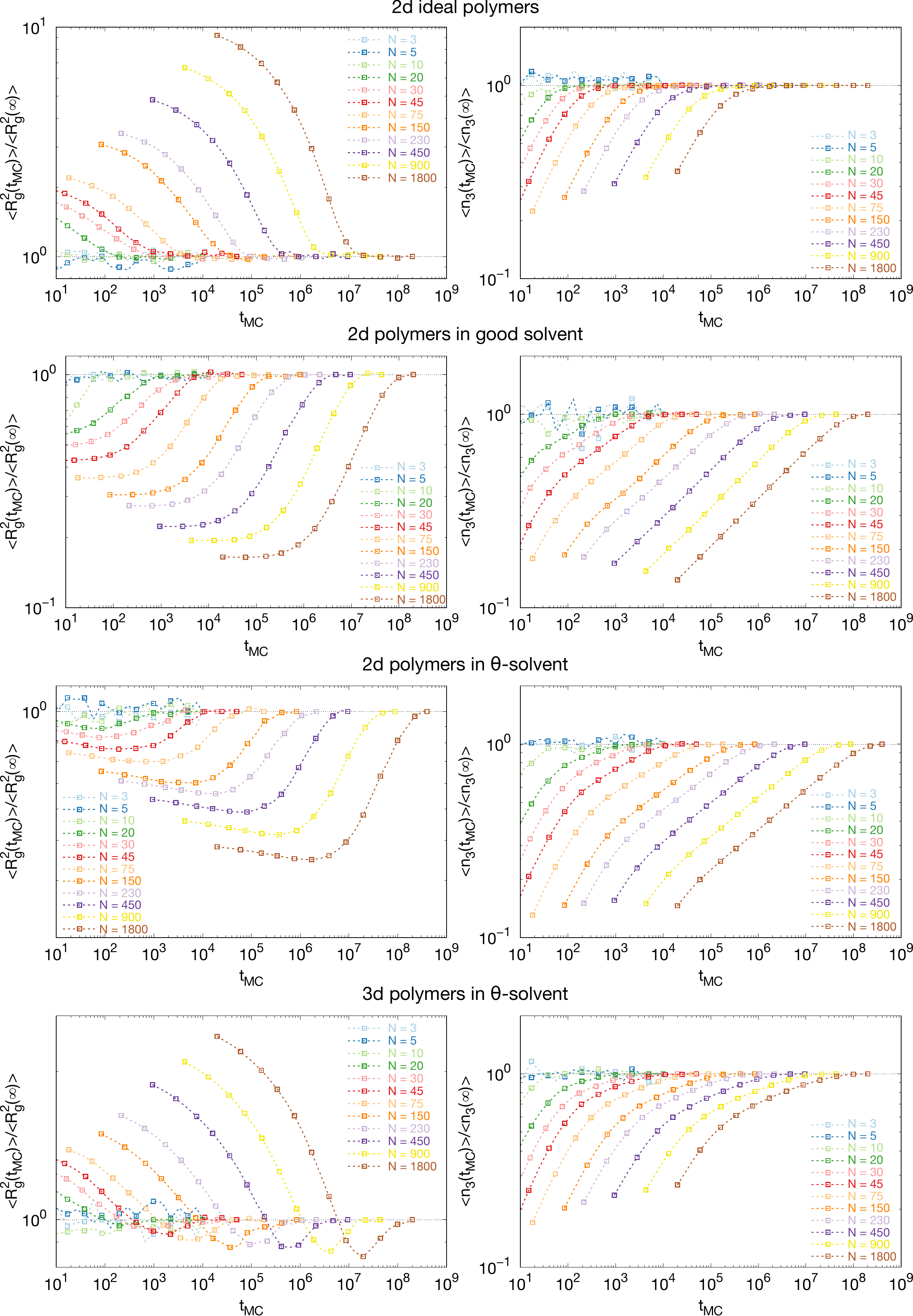}
  \caption{
  Monitoring the equilibration of $2d$ and $3d$ lattice trees of total weight $N$ in terms of Monte Carlo time steps ($\tMC$) of the ``amoeba'' algorithm (Sec.~\ref{sec:AmoebaAlgo}).
  (Left) Mean-square gyration radius.
  (Right) Mean number of branching nodes.
  Both quantities are normalized to the corresponding asymptotic values.
  }
  \label{fig:Rg2-n3-tMC}
\end{figure*}
\begin{table*}[htb!]
\caption{
Path length statistics and branching statistics I.
Expectation values and confidence intervals of corresponding observables for tree polymers of total weight $N$.
}
\begin{tabular}{cccccc}
\hline	
\hline
\\
\multicolumn{6}{c}{$2d$ ideal polymers}\\
\hline	
$N$	&	$\braket{L(N)}$	&	$\braket{\dlcenter(N)}$	&	$\braket{\dlcentermax(N)}$	&	$\braket{\Nbr(N)}$	& $\braket{n_{3}}$		\\
\hline	
3	&	1.168	$\pm$	0.006	&	0.835	$\pm$	0.012	&	1.340	$\pm$	0.047	&	0.613	$\pm$	0.072	&	0.660	$\pm$	0.048	\\
5	&	1.703	$\pm$	0.009	&	1.220	$\pm$	0.009	&	2.050	$\pm$	0.022	&	0.964	$\pm$	0.072	&	1.330	$\pm$	0.057	\\
10	&	2.754	$\pm$	0.016	&	1.917	$\pm$	0.019	&	3.300	$\pm$	0.046	&	1.581	$\pm$	0.073	&	3.370	$\pm$	0.063	\\
20	&	4.395	$\pm$	0.009	&	3.033	$\pm$	0.010	&	5.375	$\pm$	0.022	&	2.624	$\pm$	0.025	&	7.278	$\pm$	0.029	\\
30	&	5.708	$\pm$	0.014	&	3.909	$\pm$	0.014	&	7.019	$\pm$	0.030	&	3.457	$\pm$	0.026	&	11.273	$\pm$	0.035	\\
45	&	7.321	$\pm$	0.019	&	4.986	$\pm$	0.019	&	9.072	$\pm$	0.037	&	4.483	$\pm$	0.029	&	17.188	$\pm$	0.044	\\
75	&	10.025	$\pm$	0.030	&	6.820	$\pm$	0.029	&	12.535	$\pm$	0.057	&	6.219	$\pm$	0.036	&	29.071	$\pm$	0.056	\\
150	&	14.923	$\pm$	0.050	&	10.109	$\pm$	0.046	&	18.810	$\pm$	0.089	&	9.332	$\pm$	0.051	&	58.876	$\pm$	0.075	\\
230	&	19.022	$\pm$	0.067	&	12.804	$\pm$	0.061	&	24.075	$\pm$	0.113	&	11.947	$\pm$	0.064	&	90.546	$\pm$	0.096	\\
450	&	27.565	$\pm$	0.105	&	18.545	$\pm$	0.095	&	35.274	$\pm$	0.174	&	17.417	$\pm$	0.095	&	177.864	$\pm$	0.140	\\
900	&	39.822	$\pm$	0.153	&	26.709	$\pm$	0.138	&	51.232	$\pm$	0.246	&	25.165	$\pm$	0.133	&	356.388	$\pm$	0.199	\\
1800	&	57.048	$\pm$	0.156	&	38.115	$\pm$	0.137	&	73.877	$\pm$	0.246	&	36.109	$\pm$	0.134	&	713.441	$\pm$	0.189	\\
\hline
\\	
\multicolumn{6}{c}{$2d$ polymers in good solvent}\\
\hline	
$N$	&	$\braket{L(N)}$	&	$\braket{\dlcenter(N)}$	&	$\braket{\dlcentermax(N)}$	&	$\braket{\Nbr(N)}$	& $\braket{n_{3}}$		\\
\hline	
3	&	1.220	$\pm$	0.005	&	0.940	$\pm$	0.011	&	1.760	$\pm$	0.043	&	0.753	$\pm$	0.072	&	0.240	$\pm$	0.043	\\
5	&	1.813	$\pm$	0.010	&	1.328	$\pm$	0.014	&	2.360	$\pm$	0.048	&	1.094	$\pm$	0.073	&	0.670	$\pm$	0.053	\\
10	&	3.076	$\pm$	0.022	&	2.228	$\pm$	0.024	&	4.020	$\pm$	0.051	&	1.899	$\pm$	0.075	&	1.930	$\pm$	0.081	\\
20	&	5.158	$\pm$	0.014	&	3.692	$\pm$	0.014	&	6.855	$\pm$	0.029	&	3.291	$\pm$	0.027	&	4.365	$\pm$	0.035	\\
30	&	6.960	$\pm$	0.021	&	4.961	$\pm$	0.020	&	9.299	$\pm$	0.041	&	4.508	$\pm$	0.031	&	6.843	$\pm$	0.046	\\
45	&	9.317	$\pm$	0.029	&	6.595	$\pm$	0.028	&	12.441	$\pm$	0.054	&	6.090	$\pm$	0.036	&	10.513	$\pm$	0.054	\\
75	&	13.545	$\pm$	0.042	&	9.582	$\pm$	0.040	&	18.187	$\pm$	0.076	&	8.967	$\pm$	0.046	&	17.843	$\pm$	0.072	\\
150	&	22.446	$\pm$	0.073	&	15.867	$\pm$	0.067	&	30.285	$\pm$	0.132	&	15.073	$\pm$	0.072	&	36.685	$\pm$	0.097	\\
230	&	30.828	$\pm$	0.102	&	21.795	$\pm$	0.093	&	41.661	$\pm$	0.177	&	20.820	$\pm$	0.096	&	56.169	$\pm$	0.117	\\
450	&	50.551	$\pm$	0.160	&	35.703	$\pm$	0.146	&	68.730	$\pm$	0.284	&	34.426	$\pm$	0.149	&	111.058	$\pm$	0.170	\\
900	&	83.736	$\pm$	0.277	&	59.099	$\pm$	0.258	&	113.453	$\pm$	0.489	&	57.152	$\pm$	0.258	&	221.690	$\pm$	0.241	\\
1800	&	140.430	$\pm$	0.364	&	99.362	$\pm$	0.336	&	190.439	$\pm$	0.596	&	96.369	$\pm$	0.331	&	443.047	$\pm$	0.279	\\
\hline
\\
\multicolumn{6}{c}{$2d$ polymers in $\uptheta$-solvent}\\
\hline
$N$	&	$\braket{L(N)}$	&	$\braket{\dlcenter(N)}$	&	$\braket{\dlcentermax(N)}$	&	$\braket{\Nbr(N)}$	& $\braket{n_{3}}$		\\
\hline	
3	&	1.158	$\pm$	0.005	&	0.815	$\pm$	0.011	&	1.260	$\pm$	0.044	&	0.587	$\pm$	0.072	&	0.740	$\pm$	0.044	\\
5	&	1.700	$\pm$	0.008	&	1.230	$\pm$	0.009	&	2.030	$\pm$	0.017	&	0.976	$\pm$	0.072	&	1.390	$\pm$	0.055	\\
10	&	2.784	$\pm$	0.013	&	1.968	$\pm$	0.016	&	3.350	$\pm$	0.048	&	1.628	$\pm$	0.072	&	3.430	$\pm$	0.056	\\
20	&	4.512	$\pm$	0.010	&	3.150	$\pm$	0.010	&	5.649	$\pm$	0.022	&	2.755	$\pm$	0.025	&	7.316	$\pm$	0.030	\\
30	&	5.934	$\pm$	0.015	&	4.134	$\pm$	0.015	&	7.484	$\pm$	0.032	&	3.689	$\pm$	0.027	&	11.296	$\pm$	0.036	\\
45	&	7.774	$\pm$	0.022	&	5.379	$\pm$	0.021	&	9.905	$\pm$	0.044	&	4.889	$\pm$	0.031	&	17.206	$\pm$	0.043	\\
75	&	10.891	$\pm$	0.032	&	7.534	$\pm$	0.030	&	14.005	$\pm$	0.060	&	6.957	$\pm$	0.038	&	29.286	$\pm$	0.053	\\
150	&	17.429	$\pm$	0.056	&	12.100	$\pm$	0.052	&	22.771	$\pm$	0.099	&	11.374	$\pm$	0.057	&	59.142	$\pm$	0.079	\\
230	&	23.178	$\pm$	0.082	&	16.150	$\pm$	0.075	&	30.468	$\pm$	0.139	&	15.274	$\pm$	0.078	&	90.834	$\pm$	0.101	\\
450	&	37.039	$\pm$	0.149	&	25.939	$\pm$	0.137	&	49.093	$\pm$	0.239	&	24.713	$\pm$	0.134	&	177.483	$\pm$	0.166	\\
900	&	59.240	$\pm$	0.202	&	41.435	$\pm$	0.184	&	79.319	$\pm$	0.347	&	39.894	$\pm$	0.185	&	352.540	$\pm$	0.404	\\
1800	&	98.262	$\pm$	0.323	&	69.474	$\pm$	0.307	&	132.010	$\pm$	0.477	&	66.873	$\pm$	0.284	&	690.381	$\pm$	0.808	\\
\hline
\\	
\multicolumn{6}{c}{$3d$ polymers in $\uptheta$-solvent}\\
\hline	
$N$	&	$\braket{L(N)}$	&	$\braket{\dlcenter(N)}$	&	$\braket{\dlcentermax(N)}$	&	$\braket{\Nbr(N)}$	& $\braket{n_{3}}$		\\
\hline	
3	&	1.156	$\pm$	0.005	&	0.813	$\pm$	0.011	&	1.250	$\pm$	0.043	&	0.583	$\pm$	0.072	&	0.750	$\pm$	0.044	\\
5	&	1.688	$\pm$	0.008	&	1.217	$\pm$	0.008	&	2.020	$\pm$	0.014	&	0.960	$\pm$	0.071	&	1.460	$\pm$	0.054	\\
10	&	2.794	$\pm$	0.015	&	1.968	$\pm$	0.017	&	3.410	$\pm$	0.049	&	1.635	$\pm$	0.073	&	3.340	$\pm$	0.065	\\
20	&	4.470	$\pm$	0.010	&	3.105	$\pm$	0.010	&	5.540	$\pm$	0.023	&	2.704	$\pm$	0.025	&	7.340	$\pm$	0.029	\\
30	&	5.820	$\pm$	0.015	&	4.024	$\pm$	0.015	&	7.241	$\pm$	0.032	&	3.570	$\pm$	0.027	&	11.400	$\pm$	0.035	\\
45	&	7.542	$\pm$	0.021	&	5.193	$\pm$	0.020	&	9.457	$\pm$	0.041	&	4.678	$\pm$	0.030	&	17.421	$\pm$	0.042	\\
75	&	10.429	$\pm$	0.032	&	7.152	$\pm$	0.031	&	13.236	$\pm$	0.060	&	6.568	$\pm$	0.038	&	29.327	$\pm$	0.056	\\
150	&	16.038	$\pm$	0.055	&	10.988	$\pm$	0.051	&	20.600	$\pm$	0.097	&	10.228	$\pm$	0.055	&	59.350	$\pm$	0.079	\\
230	&	20.563	$\pm$	0.073	&	14.041	$\pm$	0.067	&	26.423	$\pm$	0.126	&	13.130	$\pm$	0.069	&	91.420	$\pm$	0.096	\\
450	&	30.650	$\pm$	0.109	&	20.815	$\pm$	0.100	&	39.702	$\pm$	0.182	&	19.712	$\pm$	0.100	&	179.529	$\pm$	0.138	\\
900	&	46.057	$\pm$	0.173	&	31.272	$\pm$	0.156	&	59.867	$\pm$	0.280	&	29.745	$\pm$	0.155	&	359.534	$\pm$	0.188	\\
1800	&	69.520	$\pm$	0.187	&	47.381	$\pm$	0.167	&	91.185	$\pm$	0.301	&	45.169	$\pm$	0.164	&	719.044	$\pm$	0.190	\\
\hline
\hline
\label{tab:data-scaling_rho-epsilon}
\end{tabular}
\end{table*}
\begin{table*}[htb!]
\caption{
Path length statistics and branching statistics II.
Values for the critical exponents $\rho$ and $\epsilon$, obtained from best fits of model functions with ($\Delta>0$) and without ($\Delta=0$) correction-to-scaling term (see Sec.~\ref{sec:EstimatingScalExpsAvProps}, for details) to the corresponding numerical data reported in Table~\ref{tab:data-scaling_rho-epsilon}.
Final estimates with systematic and statistical errors are highlighted in boldface.
}
\begin{tabular}{ccccc}
\hline
\hline
\\
\multicolumn{5}{c}{$2d$ ideal polymers}\\
\hline
Observable    & $\braket{L(N)}$     & $\braket{\dlcenter(N)}$ & $\braket{\dlcentermax(N)}$ & $\braket{\Nbr(N)}$  \\
\hline
$\Delta$      & $0.262\pm0.032   $ & $0.205\pm0.055   $ & $0.266\pm0.045   $ & $0.309\pm0.048$                   \\
\textsc{dof} & $6                   $ & $6                   $ & $6                   $ & $6$                    \\
$\redchi$     & $0.532               $ & $0.618               $ & $0.453               $ & $0.267$               \\
$\Q$           & $0.784               $ & $0.716               $ & $0.844               $ & $0.953$               \\
Exponent      & $\rho=0.452\pm0.015$ & $\rho=0.423\pm0.038$ & $\rho=0.458\pm0.021$ & $\epsilon=0.454\pm0.019$    \\
\hline
$\Delta$      & $0    $ & $0  $ & $0  $ & $0$                                                                    \\
\textsc{dof} & $1    $ & $1  $ & $1  $ & $1$                                                                     \\
$\redchi$     & $0.868    $ & $0.581  $ & $0.404  $ & $0.313$                                                    \\
$\Q$           & $0.351    $ & $0.446  $ & $0.525  $ & $0.576$                                                    \\
Exponent      & $\rho=0.524\pm0.002$ & $\rho=0.519\pm0.002$ & $\rho=0.533\pm0.002$ & $\epsilon=0.525\pm0.002$    \\
\hline
Average       & \multicolumn{3}{c}{$\bm{\rho = 0.485 \pm 0.042 \pm 0.038}$} & $\bm{\epsilon = 0.490 \pm 0.036 \pm 0.019}$ \\
\hline
\\
\multicolumn{5}{c}{$2d$ polymers in good solvent}\\
\hline
Observable & $\braket{L(N)}$ & $\braket{\dlcenter(N)}$ & $\braket{\dlcentermax(N)}$ & $\braket{\Nbr(N)}$ \\
\hline
$\Delta$ & $0.908\pm0.785$ & $0.383\pm0.412$ & $1.094\pm8.399$ & $0.770\pm0.523$ \\
\textsc{dof} & $4$ & $5$ & $5$ & $6$ \\
$\redchi$ & $0.776$ & $0.835$ & $1.113$ & $0.567$ \\
$\Q$ & $0.541$ & $0.525$ & $0.351$ & $0.757$ \\
Exponent & $\rho=0.738\pm0.003$ & $\rho=0.747\pm0.016$ & $\rho=0.739\pm0.003$ & $\epsilon=0.743\pm0.004$ \\
\hline
$\Delta$ & $0$ & $0$ & $0$ & $0$ \\
\textsc{dof} & $1$ & $1$ & $1$ & $1$ \\
$\redchi$ & $2.525$ & $2.325$ & $2.771$ & $2.166$ \\
$\Q$ & $0.112$ & $0.127$ & $0.096$ & $0.141$ \\
Exponent & $\rho=0.738\pm0.003$ & $\rho=0.739\pm0.004$ & $\rho=0.736\pm0.004$ & $\epsilon=0.743\pm0.004$ \\
\hline
Average & \multicolumn{3}{c}{$\bm{\rho=0.739\pm0.004\pm0.016}$} & $\bm{\epsilon=0.743\pm0.000\pm0.004}$ \\
\hline 
\\
\multicolumn{5}{c}{$2d$ polymers in $\uptheta$-solvent}\\
\hline
Observable & $\braket{L(N)}$ & $\braket{\dlcenter(N)}$ & $\braket{\dlcentermax(N)}$ & $\braket{\Nbr(N)}$ \\
\hline
$\Delta$ & $-$ & $-$ & $-$ & $-$ \\
\textsc{dof} & $-$ & $-$ & $-$ & $-$ \\
$\redchi$ & $-$ & $-$ & $-$ & $-$ \\
$\Q$ & $-$ & $-$ & $-$ & $-$ \\
Exponent & $-$ & $-$ & $-$ & $-$ \\
\hline
$\Delta$ & 0 & 0 & 0 & 0 \\
\textsc{dof} & 1 & 1 & 1 & 1 \\
$\redchi$ & 18.065 & 18.679 & 7.731 & 10.616 \\
$\Q$ & $2\cdot10^{-5}$ & $1\cdot10^{-5}$ & 0.005 & 0.001 \\
Exponent & $\rho=0.706\pm0.011$ & $\rho=0.713\pm0.015$ & $\rho=0.716\pm0.008$ & $\epsilon=0.720\pm0.011$ \\
\hline
Average & \multicolumn{3}{c}{$\bm{\rho=0.711\pm0.004\pm0.015}$} & $\bm{\epsilon=0.720 \pm 0.000 \pm 0.011}$ \\
\hline
\\
\multicolumn{5}{c}{$3d$ polymers in $\uptheta$-solvent}\\
\hline
Observable & $\braket{L(N)}$ & $\braket{\dlcenter(N)}$ & $\braket{\dlcentermax(N)}$ & $\braket{\Nbr(N)}$ \\
\hline
$\Delta$ & $0.473\pm0.088$ & $0.506\pm0.163$ & $0.463\pm0.154$ & $0.667\pm0.150$ \\
\textsc{dof} & $6$ & $6$ & $6$ & $6$ \\
$\redchi$ & $1.503$ & $1.672$ & $2.056$ & $1.081$ \\
$\Q$ & $0.173$ & $0.123$ & $0.055$ & $0.371$ \\
Exponent & $\rho=0.572\pm0.008$ & $\rho=0.576\pm0.010$ & $\rho=0.577\pm0.015$ & $\epsilon=0.584\pm0.008$ \\
\hline
$\Delta$ & $0$ & $0$ & $0$ & $0$ \\
\textsc{dof} & $1$ & $1$ & $1$ & $1$ \\
$\redchi$ & $0.263$ & $0.525$ & $0.840$ & $0.269$ \\
$\Q$ & $0.608$ & $0.469$ & $0.359$ & $0.604$ \\
Exponent & $\rho=0.591\pm0.001$ & $\rho=0.594\pm0.002$ & $\rho=0.600\pm0.003$ & $\epsilon=0.599\pm0.002$ \\
\hline
Average & \multicolumn{3}{c}{$\bm{\rho=0.585\pm0.011\pm0.015}$} & $\bm{\epsilon=0.591\pm0.007\pm0.008}$ \\
\hline
\hline
\label{tab:fit-scaling_rho-epsilon}
\end{tabular}
\end{table*}
\begin{table*}[htb!]
\caption{
Conformational statistics of lattice trees I.
Expectation values and confidence intervals of corresponding observables for tree polymers of total weight $N$.
}
\begin{tabular}{ccccc}
\hline
\hline	
\\
\multicolumn{5}{c}{$2d$ ideal polymers}\\
\hline																	
$N$	&	$\braket{\Rg^{2}(N)}$	&	$\braket{R^{2}(\braket{L})}$	&	$\Lmax$		&	$\braket{R^{2}(\Lmax)}$		\\
\hline																	
3	&	0.573	$\pm$	0.019	&	1.000	$\pm$	0.000	&	2.340	$\pm$	0.047	&	2.273	$\pm$	0.149	\\
5	&	0.899	$\pm$	0.031	&	2.122	$\pm$	0.057	&	3.670	$\pm$	0.057	&	4.030	$\pm$	0.310	\\
10	&	1.382	$\pm$	0.054	&	3.085	$\pm$	0.104	&	6.130	$\pm$	0.081	&	5.945	$\pm$	0.537	\\
20	&	2.207	$\pm$	0.029	&	4.009	$\pm$	0.038	&	10.260	$\pm$	0.042	&	10.448	$\pm$	0.282	\\
30	&	2.847	$\pm$	0.039	&	5.972	$\pm$	0.069	&	13.571	$\pm$	0.057	&	13.538	$\pm$	0.385	\\
45	&	3.610	$\pm$	0.049	&	6.936	$\pm$	0.074	&	17.647	$\pm$	0.074	&	17.532	$\pm$	0.500	\\
75	&	5.117	$\pm$	0.074	&	10.205	$\pm$	0.116	&	24.550	$\pm$	0.114	&	24.690	$\pm$	0.760	\\
150	&	7.501	$\pm$	0.105	&	14.995	$\pm$	0.166	&	37.120	$\pm$	0.178	&	37.636	$\pm$	1.133	\\
230	&	9.599	$\pm$	0.134	&	19.028	$\pm$	0.212	&	47.666	$\pm$	0.226	&	46.640	$\pm$	1.452	\\
450	&	13.794	$\pm$	0.195	&	28.014	$\pm$	0.325	&	70.060	$\pm$	0.347	&	67.656	$\pm$	2.158	\\
900	&	19.631	$\pm$	0.256	&	39.384	$\pm$	0.433	&	101.976	$\pm$	0.492	&	101.510	$\pm$	3.351	\\
1800	&	28.470	$\pm$	0.286	&	57.007	$\pm$	0.470	&	147.275	$\pm$	0.491	&	148.828	$\pm$	3.511	\\
\hline	
\\
\multicolumn{5}{c}{$2d$ polymers in good solvent}\\
\hline																	
$N$	&	$\braket{\Rg^{2}(N)}$	&	$\braket{R^{2}(\braket{L})}$	&	$\Lmax$		&	$\braket{R^{2}(\Lmax)}$		\\
\hline																	
3	&	0.776	$\pm$	0.022	&	1.000	$\pm$	0.000	&	2.760	$\pm$	0.043	&	4.120	$\pm$	0.226	\\
5	&	1.394	$\pm$	0.035	&	2.687	$\pm$	0.036	&	4.330	$\pm$	0.053	&	8.530	$\pm$	0.408	\\
10	&	2.957	$\pm$	0.083	&	5.019	$\pm$	0.070	&	7.590	$\pm$	0.094	&	18.370	$\pm$	0.966	\\
20	&	6.784	$\pm$	0.060	&	11.207	$\pm$	0.050	&	13.197	$\pm$	0.055	&	46.154	$\pm$	0.822	\\
30	&	11.118	$\pm$	0.101	&	19.262	$\pm$	0.091	&	18.098	$\pm$	0.080	&	76.408	$\pm$	1.457	\\
45	&	17.695	$\pm$	0.156	&	29.027	$\pm$	0.124	&	24.377	$\pm$	0.107	&	123.440	$\pm$	2.318	\\
75	&	34.080	$\pm$	0.304	&	61.383	$\pm$	0.290	&	35.888	$\pm$	0.151	&	249.581	$\pm$	4.470	\\
150	&	80.480	$\pm$	0.697	&	132.646	$\pm$	0.595	&	60.072	$\pm$	0.263	&	583.060	$\pm$	11.008	\\
230	&	136.353	$\pm$	1.232	&	237.233	$\pm$	1.066	&	82.826	$\pm$	0.355	&	987.330	$\pm$	18.925	\\
450	&	324.685	$\pm$	2.982	&	561.806	$\pm$	2.570	&	136.958	$\pm$	0.566	&	2429.080	$\pm$	45.923	\\
900	&	757.177	$\pm$	6.516	&	1319.930	$\pm$	5.515	&	226.413	$\pm$	0.977	&	5544.630	$\pm$	108.576	\\
1800	&	1807.139	$\pm$	10.327	&	3156.800	$\pm$	9.886	&	380.355	$\pm$	1.192	&	13151.400	$\pm$	171.613	\\
\hline
\\
\multicolumn{5}{c}{$2d$ polymers in $\uptheta$-solvent}\\
\hline																
$N$	&	$\braket{\Rg^{2}(N)}$			&	$\braket{R^{2}(\braket{L})}$			&	$\Lmax$			&	$\braket{R^{2}(\Lmax)}$			\\
\hline																	
3	&	0.474	$\pm$	0.021	&	1.000	$\pm$	0.000	&	2.260	$\pm$	0.044	&	1.580	$\pm$	0.149	\\
5	&	0.776	$\pm$	0.027	&	1.873	$\pm$	0.059	&	3.610	$\pm$	0.055	&	2.820	$\pm$	0.202	\\
10	&	1.635	$\pm$	0.053	&	3.418	$\pm$	0.084	&	6.280	$\pm$	0.062	&	7.847	$\pm$	0.562	\\
20	&	3.672	$\pm$	0.037	&	7.867	$\pm$	0.056	&	10.787	$\pm$	0.043	&	19.076	$\pm$	0.438	\\
30	&	5.887	$\pm$	0.056	&	10.960	$\pm$	0.066	&	14.458	$\pm$	0.062	&	32.424	$\pm$	0.700	\\
45	&	9.224	$\pm$	0.089	&	17.572	$\pm$	0.101	&	19.331	$\pm$	0.086	&	53.526	$\pm$	1.169	\\
75	&	16.431	$\pm$	0.152	&	30.091	$\pm$	0.167	&	27.525	$\pm$	0.120	&	103.572	$\pm$	2.146	\\
150	&	37.105	$\pm$	0.330	&	63.748	$\pm$	0.333	&	45.013	$\pm$	0.197	&	236.810	$\pm$	4.811	\\
230	&	62.200	$\pm$	0.598	&	107.936	$\pm$	0.574	&	60.446	$\pm$	0.277	&	424.070	$\pm$	8.639	\\
450	&	134.132	$\pm$	1.183	&	240.441	$\pm$	1.258	&	97.679	$\pm$	0.477	&	913.513	$\pm$	18.345	\\
900	&	310.763	$\pm$	2.848	&	547.173	$\pm$	2.810	&	158.124	$\pm$	0.694	&	2155.480	$\pm$	41.149	\\
1800	&	703.895	$\pm$	4.450	&	1310.990	$\pm$	4.915	&	263.529	$\pm$	0.953	&	5438.310	$\pm$	74.901	\\
\hline
\\	
\multicolumn{5}{c}{$3d$ polymers in $\uptheta$-solvent}\\
\hline																	
$N$	&	$\braket{\Rg^{2}(N)}$			&	$\braket{R^{2}(\braket{L})}$			&	$\Lmax$			&	$\braket{R^{2}(\Lmax)}$			\\
\hline																	
3	&	0.494	$\pm$	0.017	&	1.000	$\pm$	0.000	&	2.250	$\pm$	0.043	&	1.763	$\pm$	0.117	\\
5	&	0.756	$\pm$	0.021	&	1.797	$\pm$	0.045	&	3.540	$\pm$	0.054	&	2.925	$\pm$	0.185	\\
10	&	1.502	$\pm$	0.051	&	3.157	$\pm$	0.081	&	6.370	$\pm$	0.077	&	7.118	$\pm$	0.493	\\
20	&	2.761	$\pm$	0.026	&	4.721	$\pm$	0.027	&	10.579	$\pm$	0.042	&	13.168	$\pm$	0.272	\\
30	&	3.959	$\pm$	0.038	&	7.963	$\pm$	0.052	&	13.981	$\pm$	0.061	&	19.963	$\pm$	0.437	\\
45	&	5.591	$\pm$	0.051	&	11.715	$\pm$	0.081	&	18.417	$\pm$	0.080	&	28.601	$\pm$	0.603	\\
75	&	8.470	$\pm$	0.078	&	15.817	$\pm$	0.095	&	25.965	$\pm$	0.118	&	44.178	$\pm$	0.953	\\
150	&	15.149	$\pm$	0.143	&	29.655	$\pm$	0.191	&	40.689	$\pm$	0.194	&	80.937	$\pm$	1.787	\\
230	&	21.303	$\pm$	0.187	&	42.694	$\pm$	0.273	&	52.353	$\pm$	0.251	&	113.870	$\pm$	2.434	\\
450	&	36.335	$\pm$	0.339	&	72.887	$\pm$	0.464	&	78.895	$\pm$	0.363	&	200.131	$\pm$	4.648	\\
900	&	63.654	$\pm$	0.561	&	125.023	$\pm$	0.770	&	119.231	$\pm$	0.560	&	344.247	$\pm$	7.636	\\
1800	&	112.573	$\pm$	0.717	&	223.027	$\pm$	1.010	&	181.854	$\pm$	0.602	&	622.995	$\pm$	9.942	\\
\hline
\hline																
\label{tab:data-scaling_nu-nupath}
\end{tabular}
\end{table*}
\begin{table*}[htb!]
\caption{
Conformational statistics of lattice trees II.
Values for the critical exponents $\nu$ and $\nupath$, obtained from best fits of model functions with ($\Delta>0$) and without ($\Delta=0$) correction-to-scaling term (see Sec.~\ref{sec:EstimatingScalExpsAvProps}, for details) to the corresponding numerical data reported in Table~\ref{tab:data-scaling_nu-nupath}.
Final estimates with systematic and statistical errors are highlighted in boldface.
} 
\begin{tabular}{cccc}
\hline
\hline
\\
\multicolumn{4}{c}{$2d$ ideal polymers}\\
\hline
Observable    & $\braket{\Rg^{2}(N)}$    & $\braket{R^{2}(\braket{L})}$  & $\braket{R^{2}(\Lmax)}$ \\
\hline
$\Delta$      & $0.223\pm0.187  $ & $-$ & $-$ \\
\textsc{dof} & $6                  $ & $-$ & $-$ \\
$\redchi$     & $0.885              $ & $-$ & $-$ \\
$\Q$           & $0.505              $ & $-$ & $-$ \\
Exponent      & $\nu=0.209\pm0.063$ & $-$ & $-$ \\
\hline
$\Delta$      & $0                $ & $0                $ & $0$ \\
\textsc{dof} & $1                $ & $1                $ & $1$ \\
$\redchi$     & $0.364            $ & $2.039            $ & $0.037$ \\
$\Q$           & $0.546            $ & $0.153            $ & $0.848$ \\
Exponent      & $\nu=0.262\pm0.003$ & $\nupath=0.491\pm0.010$ & $\nupath=0.530\pm0.004$ \\
\hline
Average       & $\bm{\nu = 0.236 \pm 0.026 \pm 0.063}$ & \multicolumn{2}{c}{$\bm{\nupath = 0.510 \pm 0.020 \pm 0.010}$} \\
\hline
\\
\multicolumn{4}{c}{$2d$ polymers in good solvent}\\
\hline
Observable   & $\braket{\Rg^{2}(N)}$     & $\braket{R^{2}(\braket{L})}$   & $\braket{R^{2}(\Lmax)}$ \\
\hline
$\Delta$ & $1.072 \pm 0.542$ & $-$ & $-$ \\
\textsc{dof} & $6$ & $-$ & $-$ \\
$\redchi$ & $2.233$ & $-$ & $-$ \\
$\Q$ & $0.037$ & $-$ & $-$ \\
Exponent & $\nu=0.628 \pm 0.003$ & $-$ & $-$ \\
\hline
$\Delta$ & $0$ & $0$ & $0$ \\
\textsc{dof} & $1$ & $1$ & $1$ \\
$\redchi$ & $1.297$ & $0.093$ & $0.068$ \\
$\Q$ & $0.255$ & $0.761$ & $0.795$ \\
Exponent & $\nu=0.620 \pm 0.003$ & $\nupath=0.845 \pm 0.001$ & $\nupath=0.827 \pm 0.002$ \\
\hline
Average & $\bm{\nu=0.624 \pm 0.004 \pm 0.003}$ & \multicolumn{2}{c}{$\bm{\nupath=0.836 \pm 0.009 \pm 0.002}$} \\
\hline
\\
\multicolumn{4}{c}{$2d$ polymers in $\uptheta$-solvent}\\
\hline
Observable   & $\braket{\Rg^{2}(N)}$     & $\braket{R^{2}(\braket{L})}$   & $\braket{R^{2}(\Lmax)}$ \\
\hline
$\Delta$     & $0.166\pm0.371$ & $-$ & $0.699\pm7.988$ \\
\textsc{dof} & $6$ & $-$ & $6$ \\
$\redchi$    & $1.415$ & $-$ & $2.039$ \\
$\Q$          & $0.204$ & $-$ & $0.057$ \\
Exponent     & $\nu=0.630\pm0.102$ & $-$ & $\nupath=0.878\pm0.039$ \\
\hline
$\Delta$     & $0$ & $0$ & $0$ \\
\textsc{dof} & $1$ & $1$ & $1$ \\
$\redchi$    & $1.127$ & $0.953$ & $0.105$ \\
$\Q$          & $0.288$ & $0.329$ & $0.746$ \\
Exponent     & $\nu=0.597\pm0.003$ & $\nupath=0.869\pm0.002$ & $\nupath=0.900\pm0.003$ \\
\hline
Average      & $\bm{\nu=0.614\pm0.016\pm0.102}$ & \multicolumn{2}{c}{$\bm{\nupath=0.882\pm0.015\pm0.039}$} \\
\hline
\\
\multicolumn{4}{c}{$3d$ polymers in $\uptheta$-solvent}\\
\hline
Observable   & $\braket{\Rg^{2}(N)}$     & $\braket{R^{2}(\braket{L})}$   & $\braket{R^{2}(\Lmax)}$ \\
\hline
$\Delta$ & $1.002\pm0.268$ & $-$ & $1.054\pm0.885$ \\
\textsc{dof} & $6$ & $-$ & $4$ \\
$\redchi$ & $0.657$ & $-$ & $0.175$ \\
$\Q$ & $0.685$ & $-$ & $0.951$ \\
Exponent & $\nu=0.403\pm0.002$ & $-$ & $\nupath=0.691\pm0.015$ \\
\hline
$\Delta$ & $0$ & $0$ & $0$ \\
\textsc{dof} & $1$ & $1$ & $1$ \\
$\redchi$ & $0.203$ & $5.140$ & $0.529$ \\
$\Q$ & $0.652$ & $0.023$ & $0.467$ \\
Exponent & $\nu=0.408\pm0.001$ & $\nupath=0.685\pm0.008$ & $\nupath=0.682\pm0.009$ \\
\hline
Average & $\bm{\nu=0.405\pm0.003\pm0.002}$ & \multicolumn{2}{c}{$\bm{\nupath=0.686\pm0.004\pm0.015}$} \\
\hline
\hline
\label{tab:fit-scaling_nu-nupath}
\end{tabular}
\end{table*}
\begin{table*}[htb!]
\caption{
Effective (finite tree weight $N$) and extrapolated ($N\rightarrow\infty$) exponents $(\thetal,\tl)$.
The first were obtained by best fits of the Redner-des Cloizeaux function (Eq.~\eqref{eq:RdC-q}) to the numerical distribution functions, $\pl$, of linear paths of length $\ell$ for tree polymers of total weight $N$.
The second were estimated by best fits of the first to suitably (see Sec.~\ref{sec:EstimatingScalExpsDistrFuncts}) chosen functions with three (free $\Delta$) and two ($\Delta \equiv 1$) free parameters.
The range where each fit was performed and its statistical significance are given by, respectively, $N_{\rm min}$, $\redchi$ and $\Q$,
while final estimates with corresponding systematic and statistical errors are highlighted in boldface.
}
\begin{tabular}{ccccccc}
\hline
\hline
\\
	& & \multicolumn{2}{c}{$2d$ ideal polymers} & & \multicolumn{2}{c}{$2d$ polymers in good solvent} \\
\hline	
$N$	&	& $\thetal$	&	$\tl$	&	& $\thetal$	&	$\tl$	\\
\hline	
20	&	& 0.389	$\pm$	0.025	&	3.462	$\pm$	0.162	&	& 0.217	$\pm$	0.024	&	2.859	$\pm$	0.133	\\
30	&	& 0.447	$\pm$	0.015	&	3.188	$\pm$	0.080	&	& 0.244	$\pm$	0.014	&	2.704	$\pm$	0.073	\\
45	&	& 0.507	$\pm$	0.011	&	2.977	$\pm$	0.053	&	& 0.265	$\pm$	0.008	&	2.678	$\pm$	0.044	\\
75	&	& 0.573	$\pm$	0.008	&	2.720	$\pm$	0.031	&	& 0.277	$\pm$	0.006	&	2.622	$\pm$	0.033	\\
150	&	& 0.662	$\pm$	0.007	&	2.520	$\pm$	0.023	&	& 0.291	$\pm$	0.004	&	2.541	$\pm$	0.021	\\
230	&	& 0.704	$\pm$	0.006	&	2.460	$\pm$	0.018	&	& 0.292	$\pm$	0.004	&	2.532	$\pm$	0.019	\\
450	&	& 0.780	$\pm$	0.006	&	2.283	$\pm$	0.014	&	& 0.287	$\pm$	0.003	&	2.525	$\pm$	0.016	\\
900	&	& 0.827	$\pm$	0.004	&	2.236	$\pm$	0.009	&	& 0.295	$\pm$	0.002	&	2.522	$\pm$	0.011	\\
1800	&	& 0.870	$\pm$	0.003	&	2.188	$\pm$	0.006	&	& 0.292	$\pm$	0.002	&	2.482	$\pm$	0.008	\\
\hline
$\Nmin$		&	& $	20	$	&	$	20	$ &	& $	20	$	&	$	20	$ \\
$\Delta$	&	& $	0.408\pm0.018	$	&	$	0.495\pm0.075	$ & &	$	1.476\pm0.48	$	&	$	0.594\pm1.552	$ \\
$\redchi$	&	& $	0.423	$	&	$	1.693	$ &	& $	0.66	$	&	$	1.12	$ \\
$\Q$		&	& $	0.864	$	&	$	0.118	$ &	& $	0.682	$	&	$	0.347	$ \\
Exponent	&	& $1.017\pm0.023	$	&	$2.062\pm0.024	$ &	& $0.293\pm0.001	$	&	$2.468\pm0.022	$ \\
\hline
$\Nmin$		& &	$	450	$	&	$	230	$ &	& $	20	$	&	$	20	$ \\
$\Delta$	&	& $	1	$	&	$	1	$ &	& $	1	$	&	$	1	$ \\
$\redchi$	&	& $	6.583	$	&	$	1.603	$ &	& $	0.907	$	&	$	1.393	$ \\
$\Q$		&	& $	0.01	$	&	$	0.201	$ &	& $	0.5	$	&	$	0.203	$ \\
Exponent	&	& $0.901\pm0.004	$	&	$2.154\pm0.006	$ &	& $0.294\pm0.001	$	&	$2.492\pm0.006	$ \\
\hline		
Average		&	& $\bm{0.959\pm0.058\pm0.023}$	&	$\bm{2.108\pm0.046\pm0.024}$ &	& $\bm{0.294\pm0.001\pm0.001}$	&	$\bm{2.480\pm0.012\pm0.022}$ \\
\hline
\\
	& & \multicolumn{2}{c}{$2d$ polymers in $\uptheta$-solvent} & & \multicolumn{2}{c}{$3d$ polymers in $\uptheta$-solvent} \\
\hline	
$N$	&	& $\thetal$	&	$\tl$	&	& $\thetal$	&	$\tl$	\\
\hline	
20	&	& 0.370	$\pm$	0.031	&	3.066	$\pm$	0.165	&	& 0.369	$\pm$	0.028	&	3.254	$\pm$	0.162	\\
30	&	& 0.415	$\pm$	0.019	&	2.803	$\pm$	0.087	&	& 0.439	$\pm$	0.016	&	2.940	$\pm$	0.077	\\
45	&	& 0.439	$\pm$	0.011	&	2.691	$\pm$	0.046	&	& 0.475	$\pm$	0.011	&	2.790	$\pm$	0.046	\\
75	&	& 0.450	$\pm$	0.009	&	2.575	$\pm$	0.038	&	& 0.522	$\pm$	0.006	&	2.606	$\pm$	0.024	\\
150	&	& 0.449	$\pm$	0.008	&	2.425	$\pm$	0.028	&	& 0.570	$\pm$	0.004	&	2.431	$\pm$	0.013	\\
230	&	& 0.460	$\pm$	0.005	&	2.329	$\pm$	0.019	&	& 0.599	$\pm$	0.003	&	2.415	$\pm$	0.009	\\
450	&	& 0.442	$\pm$	0.004	&	2.250	$\pm$	0.013	&	& 0.621	$\pm$	0.002	&	2.389	$\pm$	0.006	\\
900	&	& 0.405	$\pm$	0.004	&	2.342	$\pm$	0.017	&	& 0.644	$\pm$	0.002	&	2.334	$\pm$	0.005	\\
1800	&	& 0.403	$\pm$	0.002	&	2.173	$\pm$	0.008	&	& 0.658	$\pm$	0.001	&	2.263	$\pm$	0.003	\\
\hline
$\Nmin$		&	& $	230	$	&	$	20	$ &	& $	20	$	&	$	20	$ \\
$\Delta$	&	& $	0.408\pm0.147	$	&	$	0.463\pm0.159	$ &	& $	0.562\pm0.011	$	&	$	0.285\pm0.102	$ \\
$\redchi$	&	& $	9.168	$	&	$	10.253	$ &	& $	0.665	$	&	$	12.004	$ \\
$\Q$		&	& $	0.002	$	&	$	0	$ &	& $	0.678	$	&	$	0	$ \\
Exponent	&	& $0.360\pm0.041	$	&	$2.091\pm0.035	$ &	& $0.689\pm0.005	$	&	$2.069\pm0.05	$ \\
\hline
$\Nmin$		& &	$	450	$	&	$	20	$ &	& $	450	$	&	$	450	$ \\
$\Delta$	&	& $	1	$	&	$	1	$ &	& $	1	$	&	$	1	$ \\
$\redchi$	&	& $	4.434	$	&	$	14.918	$ &	& $	0.674	$	&	$	24.532	$ \\
$\Q$		&	& $	0.035	$	&	$	0	$ &	& $	0.412	$	&	$	0	$ \\
Exponent	&	& $0.389\pm0.003	$	&	$2.206\pm0.006	$ &	& $0.671\pm0.002	$	&	$2.227\pm0.004	$ \\
\hline		
Average		& &	$\bm{0.375\pm0.015\pm0.041}$	&	$\bm{2.149\pm0.057\pm0.035}$ &	& $\bm{0.680\pm0.009\pm0.005}$	&	$\bm{2.148\pm0.079\pm0.050}$ \\
\hline
\hline
\end{tabular}
\label{tab:data-df_pl-l}
\end{table*}
\begin{table*}[htb!]
\caption{
Effective exponents $(\thetapath,\tpath)$ obtained by best fits of the Redner-des Cloizeaux function (Eq.~\eqref{eq:RdC-q}) to the numerical distribution functions, $\pr$, of end-to-end spatial vectors $\vec r$ between nodes of tree polymers of total weight $N$.
Final estimates with corresponding systematic and statistical errors are given in boldface.
}
\begin{tabular}{cccccc}
\hline
\hline
\\
 & & \multicolumn{2}{c}{$2d$ ideal polymers} & \multicolumn{2}{c}{$2d$ polymers in good solvent} \\
\hline 
$N$ & $\ell$ & $\thetapath$ & $\tpath$ & $\thetapath$ & $\tpath$ \\
\hline 
450 & 16 & $-0.078 \pm 0.008$ & $2.196 \pm 0.012$ & $1.653 \pm 0.033$ & $7.518 \pm 0.179$ \\
450 & 32 & $0.002 \pm 0.003$ & $2.020 \pm 0.005$ & $1.802 \pm 0.015$ & $7.375 \pm 0.073$ \\
900 & 16 & $-0.015 \pm 0.010$ & $2.102 \pm 0.013$ & $1.651 \pm 0.033$ & $7.594 \pm 0.183$ \\
900 & 32 & $-0.008 \pm 0.005$ & $2.057 \pm 0.007$ & $1.793 \pm 0.012$ & $7.555 \pm 0.058$ \\
1800 & 16 & $0.012 \pm 0.014$ & $2.053 \pm 0.017$ & $1.641 \pm 0.032$ & $7.558 \pm 0.177$ \\
1800 & 32 & $0.004 \pm 0.004$ & $2.021 \pm 0.007$ & $1.820 \pm 0.013$ & $7.512 \pm 0.063$ \\
1800 & 64 & $0.004 \pm 0.001$ & $2.022 \pm 0.003$ & $1.851 \pm 0.006$ & $7.515 \pm 0.028$ \\
\hline
\multicolumn{2}{c}{Average} & $\bm{-0.011 \pm 0.011 \pm 0.014}$ & $\bm{2.067 \pm 0.022 \pm 0.017}$ & $\bm{1.744 \pm 0.032 \pm 0.033}$ & $\bm{7.518 \pm 0.024 \pm 0.183}$ \\
\hline
\\
 & & \multicolumn{2}{c}{$2d$ polymers in $\uptheta$-solvent} & \multicolumn{2}{c}{$3d$ polymers in $\uptheta$-solvent} \\
\hline 
$N$ & $\ell$ & $\thetapath$ & $\tpath$ & $\thetapath$ & $\tpath$ \\
\hline 
450 & 16 & $1.067 \pm 0.019$ & $5.402 \pm 0.079$ & $0.509 \pm 0.009$ & $3.153 \pm 0.018$ \\
450 & 32 & $1.202 \pm 0.012$ & $5.929 \pm 0.054$ & $0.514 \pm 0.006$ & $2.966 \pm 0.012$ \\
900 & 16 & $1.071 \pm 0.021$ & $5.482 \pm 0.091$ & $0.505 \pm 0.015$ & $3.173 \pm 0.030$ \\
900 & 32 & $1.296 \pm 0.011$ & $5.775 \pm 0.047$ & $0.561 \pm 0.003$ & $3.069 \pm 0.006$ \\
1800 & 16 & $1.059 \pm 0.021$ & $5.527 \pm 0.095$ & $0.500 \pm 0.009$ & $3.261 \pm 0.019$ \\
1800 & 32 & $1.263 \pm 0.011$ & $5.965 \pm 0.050$ & $0.574 \pm 0.002$ & $3.096 \pm 0.005$ \\
1800 & 64 & $1.422 \pm 0.007$ & $6.198 \pm 0.029$ & $0.572 \pm 0.002$ & $3.031 \pm 0.003$ \\
\hline 
\multicolumn{2}{c}{Average} & $\bm{1.197 \pm 0.049 \pm 0.021}$ & $\bm{5.754 \pm 0.103 \pm 0.095}$ & $\bm{0.533 \pm 0.012 \pm 0.015}$ & $\bm{3.107 \pm 0.034 \pm 0.030}$ \\
\hline
\hline
\end{tabular}
\label{tab:fit-df_prl-r}
\end{table*}
\begin{table*}[htb!]
\caption{
Effective (finite tree weight $N$) and extrapolated ($N\rightarrow\infty$) exponents $(\thetatree,\ttree)$.
The first were obtained by best fits of the Redner-des Cloizeaux function (Eq.~\eqref{eq:RdC-q}) to the numerical distribution functions, $\pr$, of end-to-end spatial vectors $\vec r$ between nodes of tree polymers of total weight $N$.
As for the second we adopted the same methodology as for exponents $(\thetal,\tl)$, and the notation of the table is the same as in Table~\ref{tab:data-df_pl-l}.
In the case of $2d$ polymers in good solvent, the 3-parameter fit for $\ttree$ fails, then errors are calculated based on the ones from ideal trees for which (statistical error for 3-parameter fit)$\approx2.5$(statistical error of 2-parameter fit)$\approx4.8$(systematic error).
}
\begin{tabular}{ccccc}
\hline
\hline	
\\
	&	\multicolumn{2}{c}{$2d$ ideal polymers} & \multicolumn{2}{c}{$2d$ polymers in good solvent} \\
\hline	
$N$	&	$\thetatree$	&	$\ttree$	&	$\thetatree$	&	$\ttree$	\\
\hline	
20	&	0.368	$\pm$	0.124	&	1.128	$\pm$	0.057	&	-0.218	$\pm$	0.078	&	1.512	$\pm$	0.073	\\
30	&	0.544	$\pm$	0.128	&	1.048	$\pm$	0.045	&	-0.364	$\pm$	0.034	&	1.683	$\pm$	0.047	\\
45	&	0.272	$\pm$	0.078	&	1.173	$\pm$	0.034	&	-0.428	$\pm$	0.015	&	1.870	$\pm$	0.030	\\
75	&	0.089	$\pm$	0.037	&	1.245	$\pm$	0.020	&	-0.453	$\pm$	0.006	&	1.923	$\pm$	0.015	\\
150	&	-0.006	$\pm$	0.016	&	1.334	$\pm$	0.011	&	-0.457	$\pm$	0.002	&	1.991	$\pm$	0.007	\\
230	&	-0.065	$\pm$	0.008	&	1.398	$\pm$	0.007	&	-0.455	$\pm$	0.001	&	2.030	$\pm$	0.004	\\
450	&	-0.006	$\pm$	0.006	&	1.353	$\pm$	0.005	&	-0.454	$\pm$	0.001	&	1.997	$\pm$	0.003	\\
900	&	-0.001	$\pm$	0.004	&	1.361	$\pm$	0.004	&	-0.455	$\pm$	0.001	&	2.085	$\pm$	0.002	\\
1800	&	0.013	$\pm$	0.002	&	1.343	$\pm$	0.002	&	-0.457	$\pm$	0.000	&	2.113	$\pm$	0.001	\\
\hline
$\Nmin$   & $ 150 $ & $ 230 $ & $ 20 $ & $ - $ \\
$\Delta$  & $ 0.278 \pm 3.133 $ & $ 1.175 \pm 0.106 $ & $ 1.626 \pm 0.259 $ & $ - $ \\
$\redchi$ & $ 11.770 $ & $ 13.183 $ & $ 4.167 $ & $ - $ \\
$\Q$      & $ 7.7\cdot10^{-6} $ & $ 2.8\cdot10^{-4} $ & $ 3.4\cdot10^{-4} $ & $ - $ \\
Exponent  & $ 0.079 \pm 0.103 $ & $ 1.339 \pm 0.006 $ & $ -0.4560 \pm 0.0003 $ & $ - $ \\
\hline
$\Nmin$   & $ 230 $ & $ 230 $ & $ 450 $ & $ 450 $ \\
$\Delta$  & $ 1 $ & $ 1 $ & $ 1 $ & $ 1 $ \\
$\redchi$ & $ 2.709 $ & $ 6.628 $ & $ 2.321 $ & $ 15.579 $ \\
$\Q$      & $ 0.067 $ & $ 0.001 $ & $ 0.128 $ & $ 7.9\cdot10^{-5} $ \\
Exponent  & $ 0.024 \pm 0.003 $ & $ 1.337 \pm 0.002 $ & $ -0.4574 \pm 0.0005 $ & $ 2.154 \pm 0.002 $ \\
\hline
Average   & $\bm{0.051 \pm 0.028 \pm 0.103}$ & $\bm{1.338 \pm 0.001 \pm 0.006}$ & $\bm{-0.4567 \pm 0.0007 \pm 0.0005}$ & $\bm{2.154 \pm 0.010 \pm 0.005}$ \\
\hline
\\
	&	\multicolumn{2}{c}{$2d$ polymers in $\uptheta$-solvent} & \multicolumn{2}{c}{$3d$ polymers in $\uptheta$-solvent} \\
\hline	
$N$	&	$\thetatree$	&	$\ttree$	&	$\thetatree$	&	$\ttree$	\\
\hline	
20	&	-0.181	$\pm$	0.081	&	1.528	$\pm$	0.063	&	0.156	$\pm$	0.340	&	0.977	$\pm$	0.084	\\
30	&	-0.309	$\pm$	0.054	&	1.695	$\pm$	0.061	&	-0.766	$\pm$	0.112	&	1.462	$\pm$	0.069	\\
45	&	-0.350	$\pm$	0.027	&	1.786	$\pm$	0.039	&	-0.904	$\pm$	0.068	&	1.663	$\pm$	0.063	\\
75	&	-0.390	$\pm$	0.010	&	1.936	$\pm$	0.021	&	-0.924	$\pm$	0.032	&	1.861	$\pm$	0.043	\\
150	&	-0.379	$\pm$	0.004	&	1.934	$\pm$	0.009	&	-0.833	$\pm$	0.015	&	1.848	$\pm$	0.023	\\
230	&	-0.360	$\pm$	0.002	&	1.868	$\pm$	0.005	&	-0.798	$\pm$	0.009	&	1.878	$\pm$	0.016	\\
450	&	-0.371	$\pm$	0.001	&	1.960	$\pm$	0.004	&	-0.732	$\pm$	0.005	&	1.848	$\pm$	0.009	\\
900	&	-0.372	$\pm$	0.001	&	1.934	$\pm$	0.003	&	-0.686	$\pm$	0.003	&	1.814	$\pm$	0.006	\\
1800	&	-0.376	$\pm$	0.001	&	1.889	$\pm$	0.002	&	-0.657	$\pm$	0.002	&	1.775	$\pm$	0.003	\\
\hline
$\Nmin$   & $ 230 $ & $ 230 $ & $ 150 $ & $ 30 $ \\
$\Delta$  & $ 0.895 \pm 0.344 $ & $ 7.175 \pm 2.844 $ & $ 0.505 \pm 0.041 $ & $ 1.531 \pm 0.036 $ \\
$\redchi$ & $ 6.412 $ & $ 368.506 $ & $ 0.473 $ & $ 20.447 $ \\
$\Q$      & $ 0.011 $ & $ 0.000 $ & $ 0.623 $ & $ 1.8\cdot10^{-20} $ \\
Exponent  & $ -0.379 \pm 0.003 $ & $ 1.905 \pm 0.002 $ & $ -0.582 \pm 0.025 $ & $ 1.788 \pm 0.003 $ \\
\hline
$\Nmin$   & $ 230 $ & $ 450 $ & $ 450 $ & $ 450 $ \\
$\Delta$  & $ 1 $ & $ 1 $ & $ 1 $ & $ 1 $ \\
$\redchi$ & $ 3.228 $ & $ 30.778 $ & $ 1.583 $ & $ 5.065 $ \\
$\Q$      & $ 0.040 $ & $ 2.9\cdot10^{-8} $ & $ 0.208 $ & $ 0.024 $ \\
Exponent  & $ -0.378 \pm 0.001 $ & $ 1.867 \pm 0.003 $ & $ -0.631 \pm 0.003 $ & $ 1.750 \pm 0.005 $ \\
\hline
Average   & $\bm{-0.3787 \pm 0.0003 \pm 0.0029}$ & $\bm{1.886 \pm 0.019 \pm 0.003}$ & $\bm{-0.607 \pm 0.024 \pm 0.025}$ & $\bm{1.769 \pm 0.019 \pm 0.005}$ \\
\hline
\hline
\end{tabular}
\label{tab:data-df_pr-r}
\end{table*}
%

%
%

\clearpage
%

%

\end{document}